\documentclass[article,12pt]{IEEEtran}
\onecolumn
\IEEEoverridecommandlockouts

\usepackage{xfrac}
 
\usepackage{macros}

\allowdisplaybreaks

\begin{document}

\title{Contraction of Private Quantum Channels and Private Quantum Hypothesis Testing}
 \author{Theshani Nuradha and Mark M. Wilde \\
 \textit{\small{School of Electrical and Computer Engineering, Cornell University,
 Ithaca, New York 14850, USA}}}

\maketitle

\begin{abstract}
    A quantum generalized divergence by definition satisfies the data-processing inequality; as such, the relative decrease in such a divergence  under the action of a quantum channel is at most one. This relative decrease is formally known as the contraction coefficient of the channel and the divergence. Interestingly, there exist combinations of channels and divergences for which the contraction coefficient is strictly less than one. Furthermore, understanding the contraction coefficient is fundamental for the study of statistical tasks under privacy constraints. To this end, here we establish upper bounds on contraction coefficients for the hockey-stick divergence under privacy constraints, where privacy is quantified with respect to the quantum local differential privacy (QLDP) framework, and we fully characterize the contraction coefficient for the trace distance under privacy constraints.  Using the machinery developed, we also determine an upper bound on the contraction of both the Bures distance and quantum relative entropy relative to the normalized trace distance, under QLDP constraints. 
    Next, we apply our findings to establish bounds on the sample complexity of quantum 
    hypothesis testing under privacy constraints. Furthermore, we study various scenarios in which the sample complexity bounds are tight, while providing order-optimal quantum channels that achieve those bounds. Lastly, we show how private quantum channels provide
    fairness and Holevo information stability in quantum learning settings. 
\end{abstract}

\tableofcontents

\section{Introduction}
The increase of personal data shared online, coupled with the advancement of data extraction methods, presents significant privacy risks. Statistical privacy models aim to tackle these risks in a principled manner~\cite{DMNS06,DR14,KM14, CY16, nuradha2022pufferfishJ}. Local differential privacy (LDP) is one such model (\cref{def: LDP}), where individual data is protected while answering aggregate queries~\cite{erlingsson2014rappor}. Studying statistical problems under local privacy constraints is vital for understanding the price that we have to pay to ensure privacy. 
To this end, the contraction of statistical measures and divergences under privacy constraints is an important technical tool.  One  such tool is the contraction coefficient of the total-variation distance $ \mathrm{TV}(p,q) \coloneqq \frac{1}{2}\left\| p-q\right\|_1 $ for two probability distributions $p$ and $q$: In~\cite{kairouz2014extremal}, it was shown that, under $\varepsilon$-LDP privacy constraints,
\begin{equation}
    \sup_{\substack{M \in \cB^\varepsilon_c,\\ \mathrm{TV}(p,q) \neq 0 } } \frac{\mathrm{TV}(M(p),M(q) )}{\mathrm{TV}(p,q)} = \frac{e^\varepsilon -1}{e^\varepsilon +1} = \tanh\!\left(\sfrac{\varepsilon}{2}\right) ,
\end{equation}
%\mmw{use shorthand for this function}
where $\cB^\varepsilon_c$ denotes all $\varepsilon$-LDP mechanisms and $M(p)$ and $M(q)$ represent the probability distributions resulting from processing $p$ and $q$  by the private mechanism $M$, respectively. Bounds for the contraction coefficients of other divergences including chi-square, Hellinger, and Kullback–Leibler divergence, which are obtained by replacing total-variation distance by the respective divergence, have been studied in~\cite{DJW13,duchi2018minimax,Contraction_local_new24}. 

With the development of quantum technologies and generation of quantum data, ensuring privacy of quantum systems is an important research direction. To this end,
statistical privacy frameworks for quantum data have been developed recently, which are generalizations of classical statistical privacy frameworks~\cite{QDP_computation17, aaronson2019gentle,hirche2023quantum, nuradha2023quantum}.
Quantum local differential privacy (QLDP), which is a generalization of LDP, ensures that two distinct quantum 
states passed through a quantum channel are hard to distinguish by a measurement~\cite{hirche2023quantum}. Several works have explored applications of these frameworks in quantum machine learning~\cite{quek2021private,du2021quantum,QML_DP17, measurementQLDP22,QDP_LASSO22,QML_DPwatkins23,huang2023certified}. 

Quantifying the contraction of quantum divergences under quantum privacy constraints enables the study of statistical tasks under privacy constraints (QLDP constraints). Earlier efforts in this direction include the following: entropic inequalities under QLDP constraints on algorithms and measurements were discussed in~\cite{measurementQLDP22}, while upper bounds on contraction of divergences under QLDP algorithms were studied in~\cite{hirche2023quantum, hirche2023quantum_2}. However, finding an exact characterization of contraction coefficients of quantum divergences under privacy constraints is still largely unexplored.

One can reexamine classical and quantum information processing tasks under privacy constraints. One such task is 
symmetric hypothesis testing, which is a well studied operational task both in classical and quantum information theory~\cite{helstrom1967detection, holevo1973statistical,bae2015quantum}.
In this task, a weighted average of the type~I and type~II error probabilities is minimized, where weightings are based on prior beliefs of the occurrence of the two hypotheses. For distinguishing two probability distributions (classical) and two quantum states (quantum), asymptotically optimal error exponents are characterized by the classical Chernoff bound~\cite{chernoff1952measure,hoeffding1965asymptotically} and its quantum generalization~\cite{nussbaum_Chernoff_lower,audenaert2007discriminating, audenaert2008asymptotic}, respectively.

 In the classical setting, the non-asymptotic regime of symmetric hypothesis testing has been well studied. To this end, the number of samples (i.e., the sample complexity) needed to distinguish  two probability distributions $p$ and $q$ up to a fixed non-zero error is characterized by $\Theta\big(1/H^2(p,q)\big)$,
where 
$H^2(\cdot, \cdot)$ is the square of the Hellinger divergence, defined as 
\begin{equation}
H^2(p,q) \coloneqq 2 \left(1- \sum_{x\in \cX} \sqrt{p(x) q(x)} \right),
\end{equation}
for $p,q \in \cP(\cX)$ discrete probability distributions over the domain $\cX$ (see also the recent work~\cite{Ankit2024sample}). The sample complexity of symmetric quantum hypothesis testing was recently studied in~\cite{cheng2024sample}, showing that it is characterized by $\Theta\big(1/\left[d_{\mathrm{B}}(\rho,\sigma)\right]^2\big)$, where $\left[d_{\mathrm{B}}(\rho,\sigma)\right]^2$ is the Bures distance defined in~\eqref{eq:Bures} (see also \cref{thm:sample_C_no_private}).

Symmetric hypothesis testing under LDP constraints (when one has access to privatized data samples in contrast to the data from the original distributions) has been extensively studied in a chain of works~\cite{DJW13,duchi2018minimax,structure_optimal_16, Contraction_local_new24,Ankit2023simple}. In~\cite{DJW13}, a contraction bound for the Kullback–Leibler divergence relative to the total-variation distance has been derived, which implies that  one needs 
\begin{equation}
    n= \Theta\!\left( \frac{1}{\varepsilon^2 [\mathrm{TV}(p,q)]^2}\right)
\end{equation}
samples to achieve a constant error probability while ensuring $\varepsilon$-LDP, for $\varepsilon \ll 1$.
%where total-variation distance $\mathrm{TV}(\cdot, \cdot)$ is given by $ \mathrm{TV}(p,q) \coloneqq \| p-q\|_1 /2$.
In \cite[Lemma~2]{Contraction_local_new24}, the following lower and upper bounds were established  to achieve $\varepsilon$-LDP, for $\varepsilon >0$, and a fixed error probability of 1/10, with the prior probabilities of the distributions occurring fixed to $1/2$:
\begin{equation}
 \frac{4}{35} \max \left\{   \frac{e^\varepsilon}{2(e^\varepsilon-1)^2[\mathrm{TV}(p,q)]^2}, \frac{(e^\varepsilon+1)^2}{(e^\varepsilon-1)^2 H^2(p,q)}  \right \}  \leq n \leq  \frac{2 \ln(5) (e^{\varepsilon}+1)^2}{(e^\varepsilon -1)^2 [\mathrm{TV}(p,q)]^2} .
\end{equation}
In the quantum setting, under local privacy constraints, an upper bound on the error exponent in asymmetric hypothesis testing (minimizing the type~II error probability while the type~I error probability is fixed) was presented in \cite[Corollary~5.14]{hirche2023quantum_2}.
However, the study of the non-asymptotic regime of symmetric quantum hypothesis testing under privacy constraints has been left as an open research direction.

\subsection{Contributions}

In this paper, we study the contraction of quantum divergences under privacy constraints imposed by QLDP. As an application, we consider the  non-asymptotic regime of the statistical task of quantum symmetric hypothesis testing when we have access to privatized quantum states.

First, we derive an upper bound on the privatized contraction coefficient of the hockey-stick divergence (\cref{prop:contraction_coeff_HS},~\cref{Cor:Contraction_HS_gamma_0}). As a result, for the special case of $\gamma=1$, we find an upper bound on the privatized contraction coefficient of the normalized trace distance. Using this to establish the converse and mechanism that satisfies QLDP (\cref{prop: QLDP_mechanism}) for achievability,  we show that the privatized contraction coefficient for the normalized trace distance under $\varepsilon$-QLDP mechanisms is precisely equal to $(e^\varepsilon -1)/(e^\varepsilon +1)$. 
 Moreover, using the above results, we provide upper bounds on the contraction of the Bures distance and quantum relative entropy relative to normalized trace distance in \cref{prop:contraction_Bures_trace} and \cref{Cor:contraction_RE_TS}, respectively.

Together with the tools developed, we provide upper and lower bounds on the sample complexity when symmetric hypothesis testing is carried out with access to privatized quantum states  (\cref{thm:bounds_sample_C_private}). The upper and lower bounds  in \cref{thm:bounds_sample_C_private} do not match in general: the upper bound scales as $1/ [T(\rho,\sigma)]^2$ whereas the lower bound scales as $1/T(\rho,\sigma)$.
For the specific case in which the states are orthogonal, we prove that the sample complexity scales as  $\Theta\! \left(\left[(e^\varepsilon+1)/(e^\varepsilon-1) \right]^2\right)$, showcasing the cost of privacy (\cref{Cor:high_p_SC}). 
Furthermore, we consider a specific setting where the private mechanism is chosen from a special class of channels  (defined in~\eqref{eq:special_set_channels}) depending on the states used in the hypothesis testing task. There, we establish that the sample complexity is characterized by 
$\Theta \! \left(\left({1}/ \left[{ \varepsilon T(\rho, \sigma)}\right]\right)^2  \right)$ for $\varepsilon<1$. In this case, we show that a channel belonging to QLDP mechanisms proposed in \cref{prop: QLDP_mechanism} achieves order optimality. We also find that under certain low-privacy regimes (higher $\varepsilon$) the private sample complexity behaves similarly to the non-private setting (\cref{prop:SC_low_privacy}). We extend our analysis to private sample complexity bounds for multiple hypothesis testing and asymmetric hypothesis testing in~\cref{rem:multiple_hypo_private} and \cref{rem:asym_hypoth_SC_lower}, respectively.

Moreover, we also precisely characterize the privatized contraction coefficient of the trace distance under $(\varepsilon,\delta)$-QLDP privacy constraints in~\cref{thm:contraction_T_eps_delta}, via a different proof argument for the converse, when compared to the proof of~\cref{thm:contraction_coeff_TD}. To this end, we show that the privatized contraction coefficient under these privacy constraints evaluates to  $(e^\varepsilon -1 +2 \delta)/(e^\varepsilon +1)$, generalizing the finding in~\cref{thm:contraction_coeff_TD}.

Lastly, with the use of the contraction bounds derived in this work, in \cref{prop:Privacy-implying-fairness} we provide stronger characterizations of fairness through QLDP channels compared to the known result~\cite[Proposition~14]{nuradha2023quantum}. We also address the open question of providing bounds on the Holevo information after applying private channels identified in~\cite{caro2023information}, by formally deriving that those private channels satisfy Holevo information stability (see~\cref{prop:Bounds on mutual information due to QLDP}).

\textit{Note on related work}: During the development of our paper, we became aware of independent and concurrent work, now available at~\cite{Christoph2024sample}, which also considers
sample complexity of hypothesis testing under $\varepsilon$-local differential privacy constraints.

\subsection{Organization}

The rest of our paper is organized as follows. First we introduce notation and required background in \cref{Sec:Notations_Background}. 
We study  privatized contraction coefficients under local-quantum privacy constraints in \cref{Sec:contraction_QLDP}, which may be of independent interest for a wide spectrum of statistical tasks under privacy constraints. Using the tools developed in \cref{Sec:contraction_QLDP}, we present how the sample complexity of quantum symmetric hypothesis testing scales with the privacy parameters and the states of interest (see~\cref{sec:Application_sample_complexity}). In addition, we also characterize the privatized contraction coefficient for trace distance under $(\varepsilon,\delta)$-QLDP privacy constraints in~\cref{Sec:Contraction_eps_delta}, generalizing some results presented in \cref{Sec:contraction_QLDP}. Furthermore, we explore applications in ensuring fairness and stability in learning algorithms in~\cref{Sec:other_applications}. Lastly, we conclude our paper in \cref{Sec:conclusion} with a summary and some future directions to explore.

\section{Notation and Background} 
\label{Sec:Notations_Background}

\subsection{Notation}

We begin by reviewing basic concepts from quantum information theory and refer the reader to~\cite{wilde2017quantum,khatri2020principles} for more details. A (classical or quantum) system $R$ is identified with a finite-dimensional Hilbert space~$\cH_R$. We denote the set of linear operators acting on $\cH_R$ by $\cL(\cH_R)$. The support of a linear operator $X \in \cL(\cH_R)$ is defined to be the orthogonal complement of its kernel, and it is denoted by $\supp(X)$.
Let $\Tr\!\left[C \right]$ denote the trace of $C$ and let $\Tr_A \!\left[C\right]$ denote the partial trace of $C$ over the subsystem $A$. 
The trace norm  of a matrix $B$ is defined as $\left\|B\right\|_1 \coloneqq \Tr[\sqrt{B^\dagger B} ]$. For operators $A$ and $B$, the notation $A \geq B$ indicates that $A-B$ is a positive semi-definite (PSD) operator, while $A > B$ indicates that $A-B$ is a positive definite operator.

A quantum state $\rho_R\in\cL(\cH_R)$ on $R$ is a PSD, unit-trace operator acting on $\cH_R$. We denote the set of all quantum states in 
$\cL(\cH_R)$ by $\cD(\cH_R)$. 
A state $\rho_R$ of rank one is called pure, and we may choose a normalized vector $| \psi \rangle \in \cH_R$ satisfying $\rho_R= | \psi \rangle\!\langle \psi | $ in this case. Otherwise,
$\rho_R$ is called a mixed state. 
A quantum channel $\cN \colon \cL(\cH_A ) \to \cL(\cH_B)$ is a linear, 
completely positive, and trace-preserving (CPTP) map from $\cL(\cH_A)$ to $\cL(\cH_B)$. We often use the shorthand $\mathcal{N}_{A\to B}$ for such a channel, and we use the notation CPTP to denote the set of all quantum channels. We denote the Hilbert--Schmidt adjoint of $\cN$ by $\cN^\dagger$. A measurement of a quantum system $R$ is described by a
positive operator-valued measure (POVM) $\{M_y\}_{y \in \cY}$, which is defined to be a collection of PSD operators  satisfying $\sum_{y \in \cY} M_y= I_{R}$, where $\cY$ is a finite alphabet. According to the Born rule, after applying the above POVM to $\rho \in \cD(\cH_R)$, the probability of observing the outcome $y$ is given by~$\Tr\!\left[M_y \rho \right]$.

\subsection{Quantum Divergences}

Here we define quantum divergences that will be used throughout this work.
We call a distinguishability measure $\boldsymbol{D}(\cdot \Vert \cdot)$ a generalized divergence~\cite{SW12} if it satisfies the data-processing inequality; i.e., for every channel $\cN$, state $\rho$, and PSD operator~$\sigma$, 
\begin{equation}\label{eq:generalized-divergence}
    \boldsymbol{D}(\rho \Vert \sigma) \geq \boldsymbol{D}\!\left(\cN(\rho) \Vert \cN(\sigma) \right).
\end{equation}
For $\sigma$ a state, the normalized trace distance between the states $\rho$ and $\sigma$ is defined as
\begin{equation}
\label{eq:normalized-TD}
    T(\rho,\sigma)\coloneqq \frac{1}{2} \left\| \rho -\sigma\right\|_1.
\end{equation} 
It generalizes the total-variation distance between two probability distributions. For $\gamma \geq 1$,
the quantum hockey-stick divergence is defined as~\cite{SW12}
\begin{equation}\label{eq:hockey_stick}
    E_\gamma(\rho \Vert \sigma) \coloneqq \Tr\!\left[(\rho -\gamma \sigma)_{+} \right],
\end{equation}
where $\left(  A\right)  _{+}\coloneqq \sum_{i:a_{i}\geq0}a_{i}|i\rangle\!\langle i| $
for a Hermitian operator $A = \sum_{i}a_{i}|i\rangle\!\langle i|$.  For $\gamma=1$, observe that $ E_1(\rho \Vert \sigma)=T(\rho,\sigma)$.
The quantum relative entropy is defined as~\cite{umegaki1962conditional}
\begin{equation}
    D(\rho \Vert \sigma) \coloneqq \begin{cases}
        \Tr\!\left[ \rho( \ln \rho -\ln \sigma) \right] & \text{ if } \operatorname{supp}(\rho) \subseteq \operatorname{supp}(\sigma) \\
        +\infty & \text{ otherwise }
    \end{cases},
\end{equation}
the Uhlmann fidelity as~\cite{Uhl76_nourl}%
\begin{equation}\label{eq:fidelity}
F(\rho,\sigma) \coloneqq \left\| \sqrt{\rho} \sqrt{\sigma}\right\|_1^2,
\end{equation}
and the square of the Bures distance $d_{B}(\rho,\sigma)$ as \begin{equation}\label{eq:Bures}
\left[  d_{B}(\rho,\sigma)\right]  ^{2} \coloneqq 2\left(  1-\sqrt{F}(\rho
,\sigma)\right)  .
\end{equation}

\subsection{Privacy Models}

We review privacy frameworks for both the classical and quantum settings,  focusing especially on the local setting in which the privacy of individual entries and quantum states are ensured, respectively. 
Local differential privacy (LDP) allows for answering queries about aggregate quantities while protecting the individual entries without the need of a central trustworthy data curator~\cite{erlingsson2014rappor}. We note that a randomized privacy mechanism~$A$, as mentioned below, is described by a (regular) conditional probability distribution $P_{A|X}$ for its output given the data.
\begin{definition}[Classical Local Differential Privacy] \label{def: LDP}
Fix $\varepsilon \geq 0$ and $\delta \in [0,1]$. A~randomized mechanism ${A}: \cX \to \cY$ is $(\varepsilon,\delta)$-local differentially private if 
\begin{equation}
\PP\big(A(x) \in \cB \big) \leq e^{\varepsilon} \hspace{1mm} \PP\big(A(x') \in \cB\big) + \delta,\label{eq:dp_def}
\end{equation}
for all $x,x' \in \cX$ and $\cB \subseteq \cY $ measurable. We say that $A$ satisfies $\varepsilon$-LDP if it satisfies $(\varepsilon,0)$-LDP.
\end{definition}

Quantum local differential privacy (QLDP) ensures the privacy of quantum states, which are given as inputs to a  private quantum channel~\cite{hirche2023quantum,nuradha2023quantum}. 
\begin{definition}[Quantum Local Differential Privacy]\label{def:QLDP}
    Fix $\varepsilon \geq 0$ and $\delta \in [0,1]$. 
    Let $\cA$ be a quantum algorithm (viz., a quantum channel). The algorithm~$\cA$ is $ (\varepsilon, \delta)$-local differentially private if 
\begin{equation} \Tr\!\left[M \cA(\rho)\right] \leq e^\varepsilon \Tr\!\left[M \cA(\sigma)\right] + \delta,\qquad \forall  \rho, \sigma \in \cD(\cH), \quad \forall M: 0\leq M \leq I.
\label{eq:QLDP-def}
\end{equation}
We say that $\cA$ satisfies $\varepsilon$-QLDP if it satisfies $(\varepsilon,0)$-QLDP.
\end{definition}

% \medskip 
\begin{remark}[Connection to Hockey-stick Divergence]
$(\varepsilon,\delta)$-QLDP of
a mechanism $\cA$ is equivalent to the following condition:
\begin{equation}
\label{eq:equivalent_HS_QLDP_delta}
    \sup_{\rho,\sigma \in \cD(\cH)}  E_{e^\varepsilon}\!\left( \cA(\rho) \Vert \cA(\sigma) \right)  \leq \delta,
\end{equation}
 where $E_{\gamma}(\cdot \| \cdot )$ is defined in~\eqref{eq:hockey_stick} for $\gamma \geq 1$.
This was shown for $(\varepsilon,\delta)$-QLDP in~\cite[Eq.~(V.1)]{hirche2023quantum}. We provide a brief proof here for convenience:
For all states $\rho$ and $\sigma$ and for every measurement operator $M$ (i.e., satisfying $0\leq M \leq I$), the $\varepsilon$-QLDP criterion in~\eqref{eq:QLDP-def} can be arranged as $ \Tr\!\left[M \left(\cA(\rho) -e^\varepsilon \cA(\sigma) \right) \right]   \leq \delta$. Then, we have 
\begin{align}\label{eq:eps_delta_hockey}
 \sup_{\rho,\sigma \in \mathcal{D}(\mathcal{H})} E_{e^\varepsilon}\!\left( \cA(\rho) \Vert \cA(\sigma) \right) & = 
 \sup_{\rho,\sigma \in \mathcal{D}(\mathcal{H})}
 \Tr\!\left[\left( \cA(\rho) -e^\varepsilon \cA(\sigma)\right)_{+} \right]\\
 & =
 \sup_{\rho,\sigma \in \mathcal{D}(\mathcal{H})}
 \sup_{0\leq M\leq I} \Tr\!\left[M \left(\cA(\rho) -e^\varepsilon \cA(\sigma) \right) \right]  \\
 & \leq \delta,
\end{align}
concluding the claim. For the case in which $\delta=0$, $\varepsilon$-QLDP is equivalent to 
\begin{equation}
\label{eq:equivalent_HS_QLDP}
    \sup_{\rho,\sigma \in \cD(\cH)}  E_{e^\varepsilon}\!\left( \cA(\rho) \Vert \cA(\sigma) \right)  =0,
\end{equation}
since $\Tr\!\left[\left( \cA(\rho) -e^\varepsilon \cA(\sigma)\right)_{+} \right] \geq 0$ in~\eqref{eq:eps_delta_hockey}.

Furthermore, it is sufficient to consider the supremum over orthogonal pure states in~\eqref{eq:equivalent_HS_QLDP_delta}. That is, $(\varepsilon,\delta)$-QLDP is equivalent to
\begin{equation}
    \sup_{\rho,\sigma \in \cD(\cH)}  E_{e^\varepsilon}\!\left( \cA(\rho) \Vert \cA(\sigma) \right)  = \sup_{\varphi_{1}\perp\varphi_{2}}E_{e^\varepsilon}(\mathcal{A}(\varphi_{1}
)\Vert\mathcal{A}(\varphi_{2})) \leq \delta,
\end{equation}
where $\varphi_{1}$ and $\varphi_{2}$ are orthogonal pure
states.
This is implied by the proof of~\cite[Theorem~II.2]{hirche2023quantum}. We provide an alternative proof for this fact in~\cref{app:properties_hocky_stick}.
\end{remark}    

\begin{remark}[Connection to Max-Relative Entropy and Datta--Leditzky Divergence]
$\varepsilon$-QLDP is also equivalent to the following constraint:
    \begin{equation}
       \sup_{\rho,\sigma \in \cD(\cH)} D_{\max}\big( \cA(\rho) \Vert \cA(\sigma)\big)
       % \max\{ D_{\max}\big( \cA(\rho) \Vert \cA(\sigma)\big), D_{\max}\big( \cA(\sigma) \Vert \cA(\rho)\big)  \}
       \leq \varepsilon,
       \label{eq:dmax-qldp}
    \end{equation}
    as shown in ~\cite[Eq.~(193)]{nuradha2023quantum} (see also \cite[Lemma~III.2]{hirche2023quantum} with $\delta=0)$,
    where the max-relative entropy is defined as~\cite{datta2009min}
\begin{align}
D_{\max}(\rho\Vert\sigma) & \coloneqq\ln \inf\left\{  \lambda:\rho\leq\lambda
\sigma\right\} \label{eq:D-max-def} \\
& = \ln \sup_{0\psd M \psd I} \frac{\Tr\!\left[M \rho\right]}{\Tr\!\left[M \sigma\right]}.\label{eq:D-max-def_ALT}
\end{align}
 The inequality in \eqref{eq:dmax-qldp} can also be inferred from the dual formulation of max-relative entropy in~\eqref{eq:D-max-def_ALT}, together with~\cref{def:QLDP}.  

 { By adapting~\cite[Proposition~1]{nuradha2023quantum}, $(\varepsilon, \delta)$-QLDP is also equivalent to the following condition: 
 \begin{equation}
     \sup_{\rho,\sigma \in \cD(\cH)} \overline{\sD}^{\delta}\!\left(\cA(\rho)\Vert \cA(\sigma) \right) \leq \varepsilon,
 \end{equation}
where 
\begin{equation}
    \overline{\sD}^{\delta}(\rho\Vert\sigma)   \coloneqq  \ln \inf\left\{
\lambda\geq0:\operatorname{Tr}[\left(  \rho-\lambda\sigma\right)  _{+}%
]\leq\varepsilon \right\}.  
\end{equation}
 
 Furthermore, it is sufficient to consider the supremum over orthogonal pure states in~\eqref{eq:equivalent_HS_QLDP_delta}. That is, $(\varepsilon,\delta)$-QLDP is equivalent to
\begin{equation}
    \sup_{\rho,\sigma \in \cD(\cH)}   \overline{\sD}^{\delta}\!\left( \cA(\rho) \Vert \cA(\sigma) \right)  = \sup_{\varphi_{1}\perp\varphi_{2}} \overline{\sD}^{\delta}(\mathcal{A}(\varphi_{1}
)\Vert\mathcal{A}(\varphi_{2})) \leq \varepsilon,
\end{equation}
where $\varphi_{1}$ and $\varphi_{2}$ are orthogonal pure
states. The proof of this fact is in~\cref{app:properties_datta_leditzky}. For $\delta=0$, this can also be deduced from~\cite[Theorem 5.7]{guan2024optimal}.
%For $\delta=0$, this is also shown in
}
\end{remark}

{For further introductions to statistical privacy frameworks including quantum differential privacy, and quantum pufferfish privacy and their potential applications,  we refer to \cite{hirche2023quantum, nuradha2023quantum}.}

\section{Contraction under Quantum Privacy Constraints}

\label{Sec:contraction_QLDP}

In this section, we first propose a quantum mechanism  that  satisfies $\varepsilon$-QLDP, and we then proceed to evaluating the contraction of generalized divergences, with an emphasis on the hockey-stick divergence and normalized trace distance under privacy constraints, as imposed by QLDP.

Let $\cA^p_{\mathrm{Dep}}$ be the depolarizing channel
\begin{equation}
 \cA^p_{\mathrm{Dep}}(\rho) \coloneqq (1-p) \rho + \frac{p}{d} I,   
 \label{eq:depol-ch-def}
\end{equation}
 where $p \in[0,1]$ and $d$ is the dimension of the Hilbert space on which $\rho$ acts.
 {A depolarization mechanism achieving standard quantum differential privacy with a neighboring relation declared by the closeness of normalized trace distance was presented in \cite{QDP_computation17,hirche2023quantum}. Considering the worst case scenario where the normalized trace distance constraint evaluates to one, a mechanism achieving $\varepsilon$-QLDP based on a depolarizing channel can be obtained, as shown in \cite[Remark~9]{nuradha2023quantum}.}
%In \cite[Remark~9]{nuradha2023quantum}, a mechanism achieving $\varepsilon$-QLDP based on a depolarizing channel was presented. 
However, the parameters therein depend on $d$. 
%Here, we present a mechanism that employs a measurement channel along with the depolarizing channel to achieve $\varepsilon$-QLDP, where the parameters are independent of $d$.

{In \cite[Lemma~5.2]{angrisani2023unifying}, a general recipe to derive privacy mechanisms by applying a POVM  followed by a classical privacy mechanism, has been studied. The next proposition derives a QLDP mechanism using a similar technique, where a two-outcome POVM is followed by the application of a depolarizing mechanism.} In this mechanism, the parameters are independent of the dimension $d$ of the input states.
\begin{proposition}[QLDP Mechanism] \label{prop: QLDP_mechanism}
For all $\varepsilon \geq 0$ and $p\in[0,1]$, the mechanism  $\cA^p_{\mathrm{Dep}} \circ \cM$ satisfies $\varepsilon$-QLDP if 
 \begin{equation}
     p \geq \frac{2}{e^\varepsilon +1},
     \label{eq:depo-QLDP-mech}
 \end{equation}
 where, for a fixed measurement operator $M$ (satisfying $0  \leq M \leq I$) and for an input state $\omega$, the measurement channel $\cM$ is defined as
 \begin{equation}\label{eq:measurement_channel}
        \cM(\omega) \coloneqq \Tr[ M \omega] |0 \rangle\!\langle 0| + \Tr[ (I-M) \omega] |1 \rangle\!\langle 1|. 
    \end{equation} 
\end{proposition}

\begin{IEEEproof}
Fix $\rho, \sigma$, $d=2$, and $M$ such that $0\leq  M \leq I$  and consider that
\begin{align}
    & \frac{\Tr\!\left[M \cA^p_{\mathrm{Dep}}\!\left(\cM(\rho)\right)  \right]}{\Tr\!\left[M \cA^p_{\mathrm{Dep}}\!\left(\cM(\sigma)\right)  \right]} -1  \notag \\
    &=  \frac{ (1-p) \Tr\!\left[M \cM(\rho) \right] + \frac{p}{d}\Tr\!\left[M\right]}{(1-p) \Tr\!\left[M \cM(\sigma) \right] + \frac{p}{d}\Tr\!\left[M\right]} -1 \\ 
    &= \frac{ (1-p) \Tr\!\left[M \!\left( \cM(\rho) - \cM(\sigma)\right) \right]}{(1-p) \Tr\!\left[M \cM(\sigma) \right] + \frac{p}{d}\Tr\!\left[M\right]} \\
    & \leq \frac{ (1-p) {\left |\Tr\!\left[M \!\left( \cM(\rho) - \cM(\sigma)\right) \right] \right |}}{ \frac{p}{d}\Tr\!\left[M\right]} \numberthis .\label{eq:proof_mechanism}
\end{align}
Consider that
\begin{align}
\Tr\!\left[M \!\left( \cM(\rho) - \cM(\sigma)\right) \right] 
& \leq \left\|M\right\|_1 \frac{\left\| \cM(\rho)- \cM(\sigma) \right\|_1}{2} \\
& \leq \Tr[M],
\end{align}
where the last inequality follows because $\|M\|_1=\Tr[M]$, given that $M\geq 0$, and the normalized trace distance of quantum states is bounded from above by one. 

Given the above, together with~\eqref{eq:proof_mechanism}, if
\begin{equation}\label{eq:required_relation}
\varepsilon \geq \ln\!\left( 1+ \frac{d(1-p)}{p} \right),
\end{equation}
then 
\begin{equation}
  \frac{\Tr\!\left[M \cA^p_{\mathrm{Dep}}\!\left(\cM(\rho)\right)  \right]}{\Tr\!\left[M \cA^p_{\mathrm{Dep}}\!\left(\cM(\sigma)\right)  \right]}  \leq e^\varepsilon.
\end{equation}
By recalling that $d=2$, and rearranging terms in~\eqref{eq:required_relation}, we arrive at~\eqref{eq:depo-QLDP-mech}.
\end{IEEEproof}

\subsection{Contraction Coefficients under Privacy Constraints}

Let $\rho$ and $\sigma$ be quantum states. A generalized divergence by its definition satisfies data processing under every channel $\cN$; i.e., 
\begin{equation}
   \boldsymbol{D}(\rho \Vert \sigma) \geq  \boldsymbol{D} \!\left( \cN(\rho) \Vert \cN(\sigma)\right)  .
\end{equation}
For some cases, this contraction can be characterized more tightly if there exists $\eta_{\boldsymbol{D}} \in (0,1)$ such that the following inequality holds for all $\rho$ and $\sigma$:
\begin{equation}
\eta_{\boldsymbol{D}} \, \boldsymbol{D}(\rho \Vert \sigma)
   \geq  \boldsymbol{D} \!\left( \cN(\rho) \Vert \cN(\sigma)\right).
\end{equation}

Here we define a privatized contraction coefficient for a generalized divergence $\boldsymbol{D}$ %under privacy constraints 
as follows: 
\begin{equation}\label{eq:contraction_generalized_divergence}
    \eta_{\boldsymbol{D}}^\varepsilon \coloneqq \sup_{\substack{\cN \in \cB^\varepsilon, \\ \rho, \sigma \in \cD, \\ \boldsymbol{D}(\rho \Vert \sigma) \neq 0 }} \frac{\boldsymbol{D} \!\left( \cN(\rho) \Vert \cN(\sigma)\right)}{\boldsymbol{D}(\rho \Vert \sigma)},
\end{equation}
where $\mathcal{B}^\varepsilon$ corresponds to the set of all $\varepsilon$-QLDP mechanisms, defined formally as
\begin{equation} \label{eq:epsilon_private_Channel}
  \mathcal{B}^\varepsilon \coloneqq \left\{ \cN \in \operatorname{CPTP}: \sup_{\rho,\sigma \in \cD(\cH)} E_{e^\varepsilon}\!\left( \cN(\rho) \Vert \cN(\sigma) \right) =0 \right\} . 
\end{equation}
In the above, we made use of the previously stated fact that \eqref{eq:equivalent_HS_QLDP} is equivalent to $\varepsilon$-QLDP.

Contraction under private channels has been studied in~\cite{hirche2023quantum} by choosing normalized trace distance as the quantum divergence. In particular, Corollary~V.I therein states that, for a channel $\cN$ that satisfies $\varepsilon$-QLDP, 
\begin{equation}
    T\!\left(\cN(\rho), \cN(\sigma) \right) \leq \eta^\varepsilon_{{T}} T(\rho,\sigma),
\end{equation}
where 
\begin{equation}\label{eq:previously_tr_coeffi}
    \eta^\varepsilon_{{T}} \leq \frac{e^\varepsilon -1}{e^\varepsilon}.
\end{equation}
Observe that \cref{prop:contraction_coeff_HS} below improves this upper bound, and \cref{thm:contraction_coeff_TD} states that the bound from \cref{prop:contraction_coeff_HS} is in fact tight. 

In this spirit, we define the following privatized contraction coefficient for the hockey-stick divergence~$E_\gamma$:
\begin{equation}
    \eta_{E_\gamma}^\varepsilon \coloneqq \sup_{\substack{\cN \in \cB^\varepsilon, \\ \rho, \sigma \in \cD, \\ E_\gamma(\rho \Vert \sigma) \neq 0 }} \frac{E_\gamma\!\left( \cN(\rho) \Vert \cN(\sigma)\right)}{E_\gamma(\rho \Vert \sigma)}.
    \label{eq:contr-coef-hockey-stick-def}
\end{equation}
Observe that $\gamma=1$ corresponds to the normalized trace distance. 

Next we provide an upper bound on the privatized contraction coefficient defined in~\eqref{eq:contr-coef-hockey-stick-def} for $\gamma \in [1,e^\varepsilon]$. Note that $\eta_{E_\gamma}^\varepsilon = 0$ for $\gamma \geq e^\varepsilon$, as a consequence of~\eqref{eq:equivalent_HS_QLDP} and the fact that $E_\gamma$ is monotone non-increasing in~$\gamma$ \cite[Lemma~4.2]{datta2014second}. 
The privatized contraction coefficient of the hockey-stick divergence in the classical privacy setting (i.e., analogous to \cref{prop:contraction_coeff_HS}) was established in \cite[Theorem~1]{zamanlooy2023strong}, and our proof below is inspired by their proof.

\begin{theorem}
    [Privatized Contraction Coefficient of Hockey-Stick Divergence] \label{prop:contraction_coeff_HS}
    For $\gamma \in [1, e^\varepsilon]$, the privatized contraction coefficient $\eta^\varepsilon_{E_\gamma}$ satisfies the following inequality:
    \begin{equation}
     \eta^\varepsilon_{E_\gamma} \leq    \frac{e^\varepsilon -\gamma}{e^\varepsilon +1 }.
    \end{equation}    
\end{theorem}

\begin{IEEEproof}
Let $\cN$ be an arbitrary channel from $\mathcal{L}(\mathcal{H}_A)$ to $\mathcal{L}(\mathcal{H}_B)$. 
By~\cite[Theorem~II.2]{hirche2023quantum}, we have that
\begin{align}\label{eq:hockey-stick-contraction-general}
  \sup_{\substack{\rho, \sigma \in \cD }} \frac{E_\gamma\!\left( \cN(\rho) \Vert \cN(\sigma)\right)}{E_\gamma(\rho \Vert \sigma)} 
  &= \sup_{| \phi \rangle \perp  | \psi \rangle}  E_\gamma\!\left( \cN(| \phi\rangle\!\langle \phi|) \Vert \cN(| \psi \rangle\!\langle \psi|) \right),
\end{align}
where $| \phi \rangle \perp  | \psi \rangle$ denotes orthogonal state vectors. Then
consider that
\begin{align}
     \eta_{E_\gamma}^\varepsilon 
    &=\sup_{\cN \in \cB^\varepsilon} \sup_{| \phi \rangle \perp  | \psi \rangle}  E_\gamma\!\left( \cN(| \phi\rangle\!\langle \phi|) \Vert \cN(| \psi \rangle\!\langle \psi|) \right) \\ 
    & = \sup_{\substack{| \phi \rangle \perp  | \psi  \rangle ,\\ \mathcal{N}\, :\,  E_{e^\varepsilon}\!\left( \cN(| \phi\rangle\!\langle \phi|) \Vert \cN(| \psi \rangle\!\langle \psi|) \right) = 0,  \\ E_{e^\varepsilon}\!\left( \cN(| \psi\rangle\!\langle \psi|) \Vert \cN(| \phi \rangle\!\langle \phi|) \right)=0} } \mspace{-8mu}E_\gamma\!\left( \cN(| \phi\rangle\!\langle \phi|) \Vert \cN(| \psi \rangle\!\langle \psi|) \right) \\ 
    & \leq \sup_{\substack{ \rho', \sigma' \in \cD(\cH_B), d_B \geq 1, \\ E_{e^\varepsilon}\!\left( \rho' \Vert \sigma' \right) = E_{e^\varepsilon}\!\left( \sigma' \Vert \rho' \right)=0} } E_\gamma\!\left( \rho' \Vert \sigma' \right), \label{eq:1mid_way}
\end{align}
where the second equality follows from the hockey-stick divergence equivalent form of QLDP given in~\eqref{eq:equivalent_HS_QLDP}, and the last inequality because the set being optimized over is larger. Note that the case $d_B=1$ is trivial since the only one-dimensional state is the number 1, and the hockey-stick divergence is equal to zero for such one-dimensional states. 

Next, we show that we can restrict the above optimization to a qubit space $d_B=2$ as follows: consider the measurement channel 
\begin{equation}
    \cM(\omega) \coloneqq \Tr[ \Pi_{+} \omega] |0 \rangle\!\langle 0| + \Tr[(I-\Pi_{+}) \omega] |1 \rangle\!\langle 1|,
\end{equation}
where $\Pi_{+}$ is the projection onto the positive eigenspace of $\rho' -\gamma \sigma'$.
With that, one can check that, for all $\gamma \geq 1$,
\begin{equation}
    E_\gamma\!\left( \rho' \Vert \sigma' \right) = E_\gamma\!\left( \cM(\rho' )\Vert \cM( \sigma') \right).
\end{equation}
To this end, by data processing and non-negativity of the hockey-stick divergence, we arrive at 
\begin{equation}
    0 = E_{e^\varepsilon}\!\left( \rho' \Vert \sigma' \right) \geq E_{e^\varepsilon}\!\left( \cM(\rho')  \Vert \cM( \sigma') \right) \geq 0, 
\end{equation}
implying that $E_{e^\varepsilon}\!\left( \cM(\rho')  \Vert \cM( \sigma') \right) = 0$.
Proceeding with the above ideas, we can rewrite the right-hand side of~\eqref{eq:1mid_way} as 
\begin{equation}\label{eq:mid_processed}
 \sup_{\substack{ \rho', \sigma' \in \cD(\cH_B), d_B \geq 2, \\ E_{e^\varepsilon}\!\left( \cM(\rho') \Vert \cM(\sigma' )\right) = E_{e^\varepsilon}\!\left( \cM(\sigma') \Vert \cM(\rho') \right)=0} } E_\gamma\!\left( \cM(\rho') \Vert \cM(\sigma') \right).
\end{equation}
%\mmw{swap $\rho'$ and $\sigma'$ last term in subscript?}\tn{Yes, it is}
Define the state
\begin{equation}
    \omega_p \coloneqq  p |0 \rangle\!\langle 0| + (1-p)  |1 \rangle\!\langle 1| ,
\end{equation}
and observe that the output of the channel $\cM$ is a special case of this state with $p =\Tr[\Pi_{+} \omega]$.
 Then, we find an upper bound on~\eqref{eq:mid_processed}, which is 
\begin{equation}
 \sup_{\substack{ p,q \in [0,1], \\ E_{e^\varepsilon}( \omega_p \Vert \omega_q) = E_{e^\varepsilon}( \omega_q \Vert \omega_p )=0} } E_\gamma( \omega_p \Vert \omega_q ).
\end{equation}

Putting this together with the inequality in~\eqref{eq:1mid_way}, we arrive at 
\begin{align}
    \eta_{E_\gamma}^\varepsilon 
     & \leq \sup_{\substack{ p,q \in [0,1], \\ E_{e^\varepsilon}\!\left( \omega_p \Vert \omega_q \right) = E_{e^\varepsilon}\!\left( \omega_q \Vert \omega_p \right)=0} } E_\gamma\!\left( \omega_p \Vert \omega_q \right) \label{eq:Egamma-bernoulli-pf-1} \\ 
    & = \sup_{\substack{ p,q \in [0,1], \\ q \leq p \leq \min\{q e^\varepsilon, \ q e^{-\varepsilon}+1 -e^{-\varepsilon} \}}}  p -\gamma q \\
    &=\frac{e^\varepsilon - \gamma}{e^\varepsilon +1}.
    \label{eq:Egamma-bernoulli-pf-3}
\end{align}
The fact that~\eqref{eq:Egamma-bernoulli-pf-1} and~\eqref{eq:Egamma-bernoulli-pf-3} are equal was stated in the proof of \cite[Theorem~1]{zamanlooy2023strong}. Here we provide a short proof for convenience: The first equality above follows from the fact that the states $\omega_p$ and $\omega_q$ correspond to Bernoulli distributions with parameters $p$ and $q$, respectively, and the reasoning given below:
$E_{e^\varepsilon}\!\left( \omega_p \Vert \omega_q \right)=0$ is equivalent to $ p \leq e^\varepsilon q$ and $(1-p) \leq e^\varepsilon (1-q)$, and $E_{e^\varepsilon}\!\left( \omega_q \Vert \omega_p \right)=0$ is equivalent to $ q \leq e^\varepsilon p$ and $(1-q) \leq e^\varepsilon (1-p)$. Without loss of generality, suppose that $q \leq p$. Then all of these inequalities can be written in the following compact form: 
$q \leq p \leq \min\{q e^\varepsilon, \ q e^{-\varepsilon}+1 -e^{-\varepsilon} \}$.
The last equality follows by solving the linear program. Indeed, we can eliminate $p$ by noting that  $\min\{q e^\varepsilon, \ q e^{-\varepsilon}+1 -e^{-\varepsilon} \} \leq 1$ and thus we can maximize its value by picking $p=\min\{q e^\varepsilon, \ q e^{-\varepsilon}+1 -e^{-\varepsilon} \}$.
From there, observe that
$q e^\varepsilon \leq \ q e^{-\varepsilon}+1 -e^{-\varepsilon} $ is equivalent to $q \leq 1/(e^\varepsilon +1)$, and $q e^\varepsilon \geq \ q e^{-\varepsilon}+1 -e^{-\varepsilon} $ is equivalent to $q \geq 1/(e^\varepsilon +1)$.  With this, $p-\gamma q$ is supremized at $q=1/(e^\varepsilon +1)$.
\end{IEEEproof}

\bigskip
The hockey-stick divergence is defined as in~\eqref{eq:hockey_stick} for $\gamma \geq 1$. Now, similar to the classical definition of the hockey-stick divergence (see~\cite[Eq.~(12)]{zamanlooy2023strong}), we extend its definition to $\gamma \geq 0$ as follows: 
\begin{equation}\label{eq:hockey_stick_all_gamma}
    E_\gamma(\rho \Vert \sigma) \coloneqq \sup_{0 \leq M \leq I} \Tr\!\left[M(\rho -\gamma \sigma) \right] - (1-\gamma)_+,
\end{equation}
where $\rho$ and $\sigma$ are states and $(x)_+  \coloneqq \max\{0,x\}$.
This definition recovers~\eqref{eq:hockey_stick} because $(1-\gamma)_+=0$ for $\gamma \geq 1$.
\begin{lemma}\label{lem:hockey_stick_gamma_leq}
    Let $\gamma \geq 0$ and $\rho$ and $\sigma$ be states. Then, 
    \begin{equation}
        E_\gamma(\rho \Vert \sigma) = \gamma E_{\frac{1}{\gamma}}(\sigma \Vert \rho),
    \end{equation}
where $E_\gamma(\cdot \Vert \cdot)$ is defined in~\eqref{eq:hockey_stick_all_gamma}.
\end{lemma}
\begin{IEEEproof} Let $0 \leq \gamma \leq 1$.
    By applying~\eqref{eq:hockey_stick_all_gamma}, consider that 
    \begin{align}
         E_\gamma(\rho \Vert \sigma) &= \sup_{0 \leq M \leq I} \Tr\!\left[M(\rho -\gamma \sigma) \right] - (1-\gamma) \\
         &= \sup_{0 \leq M \leq I} \Tr\!\left[(I-M)(\rho -\gamma \sigma )\right] - (1-\gamma) \\
         &=  \sup_{0 \leq M \leq I} \Tr\!\left[M(\gamma \sigma -\rho) \right] \\
         &= \gamma \sup_{0 \leq M \leq I} \Tr\!\left[M \left( \sigma -\frac{1}{\gamma}\rho \right) \right] \\
         &= \gamma E_{\frac{1}{\gamma}}(\sigma \Vert \rho) \label{eq:relation_gamma_leq_1},
    \end{align}
    where the last inequality follows by applying~\eqref{eq:hockey_stick_all_gamma} with $1/\gamma \geq 1$ since $\gamma \leq 1$.

    For $\gamma' \geq 1$, substituting $\gamma =1/\gamma'$ in~\eqref{eq:relation_gamma_leq_1}, we conclude the proof.
\end{IEEEproof}

\medskip 
Recall that $\cA \in \cB^\varepsilon$ is equivalent to
\begin{equation}
%\label{eq:equivalent_HS_QLDP}
    \sup_{\rho,\sigma \in \cD(\cH)}  E_{e^\varepsilon}\!\left( \cA(\rho) \Vert \cA(\sigma) \right)  =0.
\end{equation}
Now, using~\cref{lem:hockey_stick_gamma_leq}, we also have that 
\begin{equation}
    \sup_{\rho,\sigma \in \cD(\cH)}  E_{e^{-\varepsilon}}\!\left( \cA(\rho) \Vert \cA(\sigma) \right)  =0.
\end{equation}
Furthermore, using the monotonicity of $\gamma' \mapsto  E_{\gamma'}$ for $\gamma' \geq 1$, we find the following for $\gamma \leq e^{-\varepsilon}$:
\begin{equation}
   E_{\frac{1}{\gamma}}\!\left( \cA(\rho) \Vert \cA(\sigma) \right) \leq  E_{e^{\varepsilon}}\!\left( \cA(\rho) \Vert \cA(\sigma) \right) =0.
\end{equation}
This then leads to $E_{\gamma}\!\left( \cA(\rho) \Vert \cA(\sigma) \right) =0 $ for $\gamma \leq e^{-\varepsilon}$ and
\begin{equation}
    \sup_{\rho,\sigma \in \cD(\cH)}  E_{\gamma}\!\left( \cA(\rho) \Vert \cA(\sigma) \right)  =0.
\end{equation}

Next, we quantify the contraction of the hockey-stick divergence for $\gamma \in [e^{-\varepsilon}, 1]$, similar to~\cref{prop:contraction_coeff_HS} for $\gamma \in [1, e^{\varepsilon}]$.

\begin{corollary}[Contraction of Hockey-Stick Divergence under QLDP]\label{Cor:Contraction_HS_gamma_0}
    Let $\gamma \in [e^{-\varepsilon}, 1]$, $\cA \in \cB^\varepsilon$, and $\rho$ and $\sigma$ states. Then, we have that 
    \begin{equation}
        E_\gamma\!\left( \cA(\rho) \Vert \cA(\sigma) \right) \leq \frac{\gamma e^\varepsilon -1}{\gamma(e^\varepsilon+1)} E_\gamma(\rho \Vert \sigma),
    \end{equation}
where $E_\gamma(\cdot \Vert \cdot)$ is defined in~\eqref{eq:hockey_stick_all_gamma}.
Furthermore, for $\gamma \in [e^{-\varepsilon}, 1]$, the privatized contraction coefficient of hockey-stick divergence, $\eta^\varepsilon_{E_\gamma} $, satisfies the following inequality:
\begin{equation}
     \eta^\varepsilon_{E_\gamma} \leq    \frac{\gamma e^\varepsilon -1}{\gamma(e^\varepsilon+1)}.
    \end{equation}   
\end{corollary}

\begin{IEEEproof}
    Since $\gamma \in [e^{-\varepsilon}, 1]$, it follows that $1/\gamma \in [1,e^{\varepsilon}]$. 
    By applying~\cref{lem:hockey_stick_gamma_leq}, we obtain the following:
    \begin{align}
         E_\gamma\!\left( \cA(\rho) \Vert \cA(\sigma) \right) &=  \gamma E_{\frac{1}{\gamma}}\!\left( \cA(\sigma) \Vert \cA(\rho) \right) \\ 
         & \leq \gamma \frac{e^\varepsilon -1/ \gamma}{e^\varepsilon +1} E_{\frac{1}{\gamma}}(\sigma \Vert \rho) \\
         &= \frac{\gamma e^\varepsilon -1}{\gamma(e^\varepsilon+1)} E_\gamma(\rho \Vert \sigma),
    \end{align}
where the first inequality follows from~\cref{prop:contraction_coeff_HS} with $1/\gamma \geq 1$ and the last equality by applying~\cref{lem:hockey_stick_gamma_leq} again.

By dividing both sides by $E_\gamma(\rho \Vert \sigma)$, and supremizing over all $\cA \in \cB^\varepsilon$ and all states such that $E_\gamma(\rho \Vert \sigma) \neq 0$, we arrive at the desired result.
\end{IEEEproof}

\medskip
\cref{prop:contraction_coeff_HS} and~\cref{Cor:Contraction_HS_gamma_0}
strengthen the previously known upper bound on the privatized contraction coefficient for the normalized trace distance (i.e., $ \eta^\varepsilon_{T}$ by choosing $\gamma=1$), as recalled in~\eqref{eq:previously_tr_coeffi}. Next, \cref{thm:contraction_coeff_TD} states that the upper bound from~\cref{prop:contraction_coeff_HS}, for trace distance, is indeed tight. Under $\varepsilon$-classical local differential privacy (recall \cref{def: LDP}),~\cref{thm:contraction_coeff_TD} was presented as \cite[Corollary~2.9]{kairouz2014extremal}. Our method of proof is different from that presented for \cite[Corollary~2.9]{kairouz2014extremal}.

\begin{theorem}[Privatized Contraction Coefficient of Trace Distance] \label{thm:contraction_coeff_TD}
%\tn{First add state dependent one and then As a consequence add this}
Let $\varepsilon \geq 0$ and $\rho$ and $\sigma$ be states such that $T(\rho,\sigma) \neq 0$. We have 
 \begin{equation}
        \sup_{\cN \in \cB^\varepsilon} \frac{T \!\left( \cN(\rho), \cN(\sigma)\right)}{T (\rho, \sigma)} = \frac{e^\varepsilon -1}{e^\varepsilon +1}.
        \label{eq:state-dependent-pcc-td}
    \end{equation}
   Consequently, the privatized contraction coefficient for the trace distance under $\varepsilon$-QLDP is given by 
    \begin{equation}
        \eta^\varepsilon_{T} = \frac{e^\varepsilon -1}{e^\varepsilon +1 }.
        \label{eq:state-independent-pcc-td}
    \end{equation}
\end{theorem}
\begin{IEEEproof}
In \cref{prop: QLDP_mechanism}, choose $M= \Pi_{+}$ for the measurement channel $\cM$, which is the projection onto the positive eigenspace of $\rho- \sigma$. 
We then conclude that
\begin{equation}
    T\!\left(\cM(\rho), \cM(\sigma)\right) = T(\rho,\sigma).
\end{equation}
To this end, consider the composition $ \cA^p_{\mathrm{Dep}} \circ \cM$, 
\begin{equation} 
    T\!\left( (\cA^p_{\mathrm{Dep}} \circ\cM)(\rho),  (\cA^p_{\mathrm{Dep}} \circ\cM)(\sigma)\right) = (1-p) T(\rho,\sigma),
\end{equation}
and choose $p =2/(e^\varepsilon +1)$ to ensure that $ \cA^p_{\mathrm{Dep}} \circ \cM$ is $\varepsilon$-QLDP. 

Then, for the channel  $ \cA^p_{\mathrm{Dep}} \circ \cM$ and states $\rho$ and $\sigma$, we find that 
\begin{equation} \label{eq:trace_dis_channelP}
    \frac{ T\!\left( (\cA^p_{\mathrm{Dep}} \circ\cM)(\rho),  (\cA^p_{\mathrm{Dep}} \circ\cM)(\sigma)\right) }{T(\rho,\sigma)} = 1-p =\frac{e^\varepsilon -1}{e^\varepsilon + 1}.
\end{equation}
In addition, by the definition of $\eta_{T}^\varepsilon$, for this choice of channel and states we get the following inequality: 
\begin{equation} \label{eq:lower_th2}
     \frac{ T\!\left( (\cA^p_{\mathrm{Dep}} \circ\cM)(\rho),  (\cA^p_{\mathrm{Dep}} \circ\cM)(\sigma)\right) }{T(\rho,\sigma)}=\frac{e^\varepsilon -1}{e^\varepsilon + 1} \leq \eta_{T}^\varepsilon.
\end{equation}

By picking $\gamma=1$ in \cref{prop:contraction_coeff_HS},
we also conclude that
\begin{equation}\label{eq:upper_thm2}
   \eta_{E_1}^\varepsilon=\eta_{T}^\varepsilon \leq \frac{e^\varepsilon -1}{e^\varepsilon + 1}.
\end{equation}
We conclude the equality in \eqref{eq:state-dependent-pcc-td} by combining 
\eqref{eq:lower_th2} and~\eqref{eq:upper_thm2}, and we conclude the equality in \eqref{eq:state-independent-pcc-td}  by supremizing over $\rho$ and $\sigma$ such that $T(\rho,\sigma) \neq 0$.
\end{IEEEproof}

\bigskip
Next, we apply the findings of \cref{prop:contraction_coeff_HS} and \cref{thm:contraction_coeff_TD} to obtain contraction bounds for other quantum divergences under QLDP privacy constraints.

\begin{proposition}[Contraction of Bures Distance under $\varepsilon$-QLDP] \label{prop:contraction_Bures_trace}
    Let $\cA$ be an $\varepsilon$-QLDP mechanism, and let $\rho$ and $\sigma$ be states. Then
    \begin{equation}
        \left[d_B\!\left( \cA(\rho), \cA(\sigma) \right)\right]^2  %\min\!\left\{ 2\frac{(e^\varepsilon -1) e^{\varepsilon/2}}{(e^\varepsilon +1) (e^{\varepsilon/2} +1)}, 
        % \leq H_{1/2}\!\left(\cA(\rho) \Vert \cA(\sigma) \right)
        \leq 2 \frac{(e^{\varepsilon/2}-1)^2}{e^\varepsilon+1}  T(\rho,\sigma).
    \end{equation}
\end{proposition}

\begin{IEEEproof}
Recall from  \cite[Eq.~(5.50)]{hirche2023quantum_2}, 
\begin{equation}\label{eq:Bures_with_alpha_helinger}
\left[d_B\!\left( \rho, \sigma \right)\right]^2 
   =   2\left(1-\sqrt{F}(\rho,\sigma)\right) \leq H_{1/2}\!\left(\rho \Vert \sigma \right),
\end{equation}
where 
\begin{equation}
    H_{1/2}\!\left(\rho \Vert \sigma \right) \coloneqq \frac{1}{2} \int_{1}^\infty \left[ E_\gamma(\rho \Vert \sigma) + E_\gamma(\sigma \Vert \rho) \right] \gamma^{-3/2} \dd \gamma.
\end{equation}
Now, as a consequence of the fact that $\cA \in \cB^\varepsilon$ implies that $E_{e^\varepsilon}\!\left(\cA(\rho) \Vert \cA(\sigma) \right)=0$ and $E_{e^\varepsilon}\!\left(\cA(\sigma) \Vert \cA(\rho) \right)=0$,  consider that
\begin{equation}
\label{eq:contraction_Bures_trace1}
     \sup_{\cA \in \cB^\varepsilon} H_{1/2}\!\left(\cA(\rho) \Vert \cA(\sigma) \right) \leq  
    \sup_{\substack{\cA \, :\,  E_{e^\varepsilon}\!\left(\cA(\rho) \Vert \cA(\sigma) \right)=0, \\ E_{e^\varepsilon}\!\left(\cA(\sigma) \Vert \cA(\rho) \right)=0}}H_{1/2}\!\left(\cA(\rho) \Vert \cA(\sigma) \right) .
\end{equation}
We use this to show the following, employing similar techniques used in the proof of~\cite[Theorem~1]{Contraction_local_new24}:
\begin{equation}
  \sup_{\cA \in \cB^\varepsilon} H_{1/2}\!\left(\cA(\rho) \Vert \cA(\sigma) \right) 
  \leq 2 \frac{(e^{\varepsilon/2}-1)^2}{e^\varepsilon+1} T(\rho,\sigma).
\end{equation}
To this end, by definition,
\begin{align}
   &  \sup_{\substack{\cA \, :\,  E_{e^\varepsilon}\!\left(\cA(\rho) \Vert \cA(\sigma) \right)=0, \\ E_{e^\varepsilon}\!\left(\cA(\sigma) \Vert \cA(\rho) \right)=0}}H_{1/2}\!\left(\cA(\rho) \Vert \cA(\sigma) \right) \cr
   &= \sup_{\substack{\cA \, :\, E_{e^\varepsilon}\!\left(\cA(\rho) \Vert \cA(\sigma) \right)=0, \\ E_{e^\varepsilon}\!\left(\cA(\sigma) \Vert \cA(\rho) \right)=0}} \frac{1}{2} \int_{1}^\infty \left[ E_\gamma\!\left(\cA(\rho) \Vert \cA(\sigma) \right) + E_\gamma\!\left(\cA(\sigma) \Vert \cA(\rho) \right) \right] \gamma^{-3/2} \dd \gamma\\
   &= \sup_{\substack{\cA \, :\, E_{e^\varepsilon}\!\left(\cA(\rho) \Vert \cA(\sigma) \right)=0, \\ E_{e^\varepsilon}\!\left(\cA(\sigma) \Vert \cA(\rho) \right)=0}} \frac{1}{2} \int_{1}^{e^\varepsilon} \left[ E_\gamma\!\left(\cA(\rho) \Vert \cA(\sigma) \right) + E_\gamma\!\left(\cA(\sigma) \Vert \cA(\rho) \right) \right] \gamma^{-3/2} \dd \gamma \\
   & \leq \sup_{\substack{\cA \, :\, E_{e^\varepsilon}\!\left(\cA(\rho) \Vert \cA(\sigma) \right)=0, \\ E_{e^\varepsilon}\!\left(\cA(\sigma) \Vert \cA(\rho) \right)=0}} \frac{1}{2}\int_{1}^{e^\varepsilon} \frac{e^\varepsilon -\gamma}{e^\varepsilon +1} \left[E_\gamma\!\left(\rho \Vert \sigma \right) + E_\gamma\!\left(\sigma \Vert \rho \right) \right] \gamma^{-3/2} \dd \gamma \\
   &=\frac{1}{2}\int_{1}^{e^\varepsilon} \frac{e^\varepsilon -\gamma}{e^\varepsilon +1} \left[E_\gamma\!\left(\rho \Vert \sigma \right) + E_\gamma\!\left(\sigma \Vert \rho \right) \right] \gamma^{-3/2} \dd \gamma \\
   & \leq E_1\!\left(\rho\| \sigma \right) \int_{1}^{e^\varepsilon} \frac{e^\varepsilon -\gamma}{e^\varepsilon+1} \gamma^{-3/2} \dd \gamma   \\
   &= 2 \frac{(e^{\varepsilon/2}-1)^2}{(e^\varepsilon+1)} T(\rho,\sigma),
\end{align}
where the second equality follows from the fact that $E_{e^\varepsilon}\!\left(\cA(\rho) \Vert \cA(\sigma) \right)=E_{e^\varepsilon}\!\left(\cA(\sigma) \Vert \cA(\rho) \right)=0$ and $\gamma \mapsto E_\gamma(\cdot \Vert \cdot)$ is a monotonically decreasing function for $\gamma \geq 1$; first inequality by applying \cref{prop:contraction_coeff_HS}; third equality by $E_\gamma(\rho \Vert\sigma) \leq E_1(\rho \Vert\sigma)$ for $\gamma \geq 1$; and finally the last equality by solving the integral and substituting $E_1(\cdot \Vert \cdot)= T(\cdot,\cdot)$.

With the use of~\eqref{eq:Bures_with_alpha_helinger} and the inequality derived above, we conclude the proof of the upper bound.
\end{IEEEproof}

\begin{remark}[Another Contraction for Bures Distance]
    By \cite[Theorem~2.1]{zhang2016lower} we have
    \begin{equation}
        1- \sqrt{F}\big( \cA(\rho), \cA(\sigma) \big) \leq T\big( \cA(\rho), \cA(\sigma) \big) \frac{e^{D_{\max}( \cA(\rho) \Vert \cA(\sigma))/2}}{1+e^{D_{\max}( \cA(\rho) \Vert \cA(\sigma))/2}}.
    \end{equation}
    Applying \cref{thm:contraction_coeff_TD} together with the fact that $\cA \in \cB^\varepsilon$ is equivalent to 
    \begin{equation}
       \sup_{\rho,\sigma \in \cD(\cH)} D_{\max}\big( \cA(\rho) \Vert \cA(\sigma)\big)
       \leq \varepsilon,
    \end{equation}
 for all $\rho,\sigma \in \cD$, we obtain the following: 
 \begin{equation}
     \left[d_B\!\left( \cA(\rho), \cA(\sigma) \right)\right]^2 \leq   2\frac{(e^\varepsilon -1) e^{\varepsilon/2}}{(e^\varepsilon +1) (e^{\varepsilon/2} +1)} T(\rho,\sigma).
 \end{equation}
 However, this bound is weaker than the bound presented in \cref{prop:contraction_Bures_trace}.
\end{remark}

\begin{proposition}
    [Contraction of Quantum Relative Entropy under Local Privacy] \label{Cor:contraction_RE_TS}
    Let $\cA$ be an $\varepsilon$-QLDP mechanism, and let $\rho$ and $\sigma$ be quantum states. Then
    \begin{equation}\label{eq:RelativeRE_TD}
        D\!\left( \cA(\rho) \Vert \cA(\sigma) \right) \leq \varepsilon \left(\frac{e^\varepsilon -1}{e^\varepsilon +1}\right) T(\rho,\sigma). 
    \end{equation}
\end{proposition}

\begin{IEEEproof}
    Using the integral form of the quantum relative entropy in~\cite[Eq.~(1.6)]{hirche2023quantum_2}, we have 
    \begin{align}
         D\!\left( \cA(\rho) \Vert \cA(\sigma) \right) &= \int_1^\infty \frac{1}{\gamma} E_\gamma\!\left(\cA(\rho) \Vert \cA(\sigma)\right) +  \frac{1}{\gamma^2} E_\gamma\!\left(\cA(\sigma) \Vert \cA(\rho)\right) \dd \gamma \\
          &={\int_1^{r_1}} \frac{1}{\gamma} E_\gamma\!\left(\cA(\rho) \Vert \cA(\sigma)\right) \dd \gamma
          +{\int_1^{r_2}} \frac{1}{\gamma^2} E_\gamma\!\left(\cA(\sigma) \Vert \cA(\rho)\right) \dd \gamma,
    \end{align}
    where
    \begin{equation}
        r_1 \equiv e^{D_{\max}(\cA(\rho) \Vert \cA(\sigma))} ,\qquad r_2 \equiv e^{D_{\max}(\cA(\sigma) \Vert \cA(\rho))},
    \end{equation}
    and we made use of \cite[Eq.~(2.3)]{hirche2023quantum_2}.
Since $\cA \in \cB^\varepsilon$, we have $ \sup_{\rho,\sigma \in \cD(\cH)} D_{\max}\big( \cA(\rho) \Vert \cA(\sigma)\big)
       \leq \varepsilon$
and we arrive at 
\begin{align}
     D\!\left( \cA(\rho) \Vert \cA(\sigma) \right) & \leq {\int_1^{e^{\varepsilon}}} \frac{1}{\gamma} E_\gamma\!\left(\cA(\rho) \Vert \cA(\sigma)\right) \dd \gamma
          +{\int_1^{e^{\varepsilon}}} \frac{1}{\gamma^2} E_\gamma\!\left(\cA(\sigma) \Vert \cA(\rho)\right) \dd \gamma \\
          & \leq \int_{1}^{e^\varepsilon} \frac{1}{\gamma} \frac{e^\varepsilon - \gamma}{e^\varepsilon +1} E_\gamma(\rho \Vert \sigma) + \frac{1}{\gamma^2} \frac{e^\varepsilon - \gamma}{e^\varepsilon +1} E_\gamma(\sigma \Vert \rho) \dd \gamma \\ 
          & \leq T(\rho,\sigma) \left( \int_{1}^{e^\varepsilon} \frac{1}{\gamma} \frac{e^\varepsilon - \gamma}{e^\varepsilon +1}  + \frac{1}{\gamma^2} \frac{e^\varepsilon - \gamma}{e^\varepsilon +1} \dd \gamma \right)  \\
          &= \varepsilon \frac{e^\varepsilon-1}{e^\varepsilon +1}T(\rho,\sigma) ,
\end{align}
where the second inequality follows from \cref{prop:contraction_coeff_HS}; the third inequality from $E_\gamma(\rho \Vert \sigma) \leq E_1(\rho \Vert \sigma)=T(\rho,\sigma) $ for all $\gamma \geq 1$ due to monotonicity of the hockey-stick divergence with respect to $\gamma$; and finally the last equality by evaluating the integral. 
\end{IEEEproof}

\medskip

Note that the inequality in~\eqref{eq:RelativeRE_TD} can alternatively be obtained by using \cite[Proposition~5.3]{hirche2023quantum_2} together with \cref{thm:contraction_coeff_TD} obtained in this work.  In that case, it is also important to note that \cref{prop:contraction_coeff_HS} is a key ingredient in the proof of \cref{thm:contraction_coeff_TD} to obtain a tight privatized contraction coefficient for the trace distance.

\medskip
Let $f: (0,\infty) \to \RR$ be a convex and twice differentiable function 
satisfying $f(1)=0$. Then, for all quantum states $\rho$ and $\sigma$, recall the quantum $f$-divergence  defined in \cite[Definition~2.3]{hirche2023quantum_2}: 
\begin{equation}\label{eq:f_divergence}
    D_f(\rho \Vert \sigma) \coloneqq \int_{1}^{\infty} f''(\gamma) E_\gamma(\rho \Vert \sigma) + \gamma^{-3} f''(\gamma^{-1}) E_\gamma(\sigma \Vert \rho) \dd \gamma.
\end{equation}
Using~\cref{lem:hockey_stick_gamma_leq}, we obtain the following equivalent expression for the above quantum $f$-divergence.
\begin{proposition}[Equivalent Expression for $f$-Divergence]
    Let $\rho$ and $\sigma$ be states. Then, $D_f(\rho \Vert \sigma)$ defined in~\eqref{eq:f_divergence} is equivalent to the following expression:
    \begin{equation}
         D_f(\rho \Vert \sigma)= \int_{0}^{\infty} f''(\gamma) E_\gamma(\rho \Vert \sigma) \dd \gamma,
    \end{equation}
where $E_\gamma(\cdot \Vert \cdot)$ is defined in~\eqref{eq:hockey_stick_all_gamma}.
\end{proposition}
\begin{IEEEproof}
    Consider that 
\begin{align}
    \int_{1}^{\infty} \gamma^{-3} f''(\gamma^{-1}) E_\gamma(\sigma \Vert \rho) \dd \gamma &= \int_{1}^{\infty} \gamma^{-3} f''(\gamma^{-1}) \gamma E_{\frac{1}{\gamma}}(\rho \Vert \sigma) \dd \gamma \\ 
    &= \int_{1}^{\infty} \gamma^{-2} f''(\gamma^{-1}) E_{\frac{1}{\gamma}}(\rho \Vert \sigma) \dd \gamma \\
    &=\int_{0}^{1}  f''(\nu) E_{\nu}(\rho \Vert \sigma) \dd \nu
\end{align}
where the first equality follows from~\cref{lem:hockey_stick_gamma_leq} and the last equality follows by change of variables with the substitution $\nu=\gamma^{-1}$.

Then, we have that 
\begin{align}
     D_f(\rho \Vert \sigma) &= \int_{1}^{\infty} f''(\gamma) E_\gamma(\rho \Vert \sigma) \dd \gamma + \int_{0}^{1}  f''(\nu) E_{\nu}(\rho \Vert \sigma) \dd \nu \\
     &=\int_{0}^{\infty}  f''(\gamma) E_{\gamma}(\rho \Vert \sigma) \dd \gamma,
\end{align}
concluding the proof.
\end{IEEEproof}

\medskip
Similar to previous cases, the privatized contraction of such an $f$-divergence (defined in~\eqref{eq:f_divergence}) relative to the normalized trace distance can be bounded from above as follows:

\begin{proposition}[Contraction of $f$-Divergences under Privacy Constraints] Let $\cA$ be an $\varepsilon$-QDP mechanism. Then for all $\rho,\sigma$  states we have
\begin{equation}\label{eq:contraction_f_d}
      D_f\!\left(\cA(\rho) \Vert \cA(\sigma) \right) \leq \frac{f\!\left( e^\varepsilon\right) + e^\varepsilon f\!\left( e^{-\varepsilon}\right) }{e^\varepsilon +1} T(\rho,\sigma),
\end{equation}  
where $D_f(\cdot \Vert \cdot)$ is defined in~\eqref{eq:f_divergence}.
\end{proposition}
\begin{IEEEproof}
    Consider that \begin{align}
    D_f\!\left(\cA(\rho) \Vert \cA(\sigma) \right) & \leq \int_{1}^{e^\varepsilon} f''(\gamma) \frac{e^\varepsilon - \gamma} {e^\varepsilon +1} E_\gamma(\rho \Vert \sigma) + \gamma^{-3} f''(\gamma^{-1}) \frac{e^\varepsilon - \gamma} {e^\varepsilon +1} E_\gamma(\sigma \Vert \rho) \dd \gamma\\ 
    & \leq \frac{1}{e^\varepsilon +1} \left( \int_{1}^{e^\varepsilon} (f''(\gamma)   + \gamma^{-3} f''(\gamma^{-1})) (e^\varepsilon -\gamma) \dd \gamma \right) T(\rho,\sigma) \\
    &= \frac{f\!\left( e^\varepsilon\right) + e^\varepsilon f\!\left( e^{-\varepsilon}\right) }{e^\varepsilon +1} T(\rho,\sigma),
\end{align}
where the first inequality follows from \cref{prop:contraction_coeff_HS} together with~\eqref{eq:equivalent_HS_QLDP}, the second inequality from $E_\gamma(\rho \Vert \sigma) \leq T(\rho,\sigma)$ for all $\gamma \geq 1$ and the last inequality by integration by parts with the use of $f(1)=0$ and $\frac{\dd^2}{\dd \gamma^2} \gamma f(\gamma^{-1})= \gamma^{-3} f''(\gamma^{-1})$ (also see the proof of~\cite[Proposition~5.2]{hirche2023quantum_2}).  
\end{IEEEproof}

\medskip
Note that the inequality in~\eqref{eq:contraction_f_d} is tighter than the bounds obtained by combining~\cite[Proposition~5.2]{hirche2023quantum_2} with the contraction bound previously known for trace distance (i.e., that in~\eqref{eq:previously_tr_coeffi}).

{
\begin{proposition}\label{prop:tightness_f_contraction}
    For $\varepsilon \geq 0$, the following equality holds:
    \begin{equation}
        \sup_{\substack{A \in \cB^\varepsilon,\\ \rho,\sigma \in \cD }}   D_f\!\left(\cA(\rho) \Vert \cA(\sigma) \right)= \frac{f\!\left( e^\varepsilon\right) + e^\varepsilon f\!\left( e^{-\varepsilon}\right) }{e^\varepsilon +1},
    \end{equation}
    where $D_f(\cdot \Vert \cdot)$ is defined in~\eqref{eq:f_divergence} and the optimization is over all $\varepsilon$-QLDP channels $\cA$ and states $\rho$ and $\sigma$.
The equality is achieved by a pair of orthogonal states and a channel $\cA$ with the output dimension two.
%for all pairs of orthogonal states that achieves the equality.
% where the equality is achieved by all pairs of orthogonal states when processed by a specific channel $\cA$ with the output dimension two.
\end{proposition}
\begin{IEEEproof}
    See~\cref{App:proof_prop_f_contr_tight}.
\end{IEEEproof}}

\section{Applications to Private Quantum Hypothesis Testing} \label{sec:Application_sample_complexity}

In this section, we first review the problem setup of symmetric hypothesis testing without privacy constraints, and then we formulate the private hypothesis testing setup. With the use of tools developed in \cref{Sec:contraction_QLDP}, we characterize bounds on the sample complexity of the private variant, both in the general and special settings.

\subsection{Quantum Hypothesis Testing with No Privacy Constraints} 

\label{Sec:General_SC_o_privacy}

\textit{Problem setup}: 
Suppose that there are two states $\rho$ and $\sigma$, and  $\rho^{\otimes n}$ is selected with probability $p\in (0,1)$ and $\sigma^{\otimes n}$ is selected with probability $q=1-p$. The sample complexity is equal to the minimum value of $n$ needed to reach a constant error probability in deciding which state was selected. To define this quantity formally, let us recall that the
Helstrom--Holevo theorem \cite{helstrom1967detection,holevo1973statistical} states that the optimal error probability $p_{e}(\rho
,\sigma,p,q)$ of symmetric quantum hypothesis testing is as follows:
\begin{align}
p_{e}(\rho,\sigma,p,q)  &  \coloneqq \min_{M_{1},M_{2}\geq0 } \left\{p\operatorname{Tr}
[M_{2}\rho]+q\operatorname{Tr}[M_{1}\sigma]:M_1 + M_2 = I \right\}
\\
&  =\frac{1}{2}\left(  1-\left\Vert p\rho-q\sigma\right\Vert _{1}\right) \label{eq:probability_error} .
\end{align}

With this in mind, we are assuming in this paradigm that there is a constant $ \alpha \in [0,1]$, and our goal is to determine the minimum value of $n$ such
that
\begin{equation}
p_{e}\!\left(\rho^{\otimes n},\sigma^{\otimes n},p,q \right) \coloneqq \frac{1}{2}\left(  1-\left\Vert p\rho^{\otimes n}-q\sigma^{\otimes
n}\right\Vert _{1}\right) \leq \alpha.\label{eq:eps-n-relation}%
\end{equation}
To this end, let us define 
\begin{equation}\label{eq:non_private_SC}
    \mathrm{SC}_{(\rho,\sigma)}(\alpha,p,q) \coloneqq \min\! \left\{ n\in \mathbb{N} : p_{e}\!\left(\rho^{\otimes n},\sigma^{\otimes n},p,q \right) \leq  \alpha \right\}. 
\end{equation}
We also use the shorthand $\mathrm{SC}_{(\rho,\sigma)} $ to refer to this quantity when $(\alpha,p,q)$ are clear from the context.

\begin{theorem}[Sample Complexity of Symmetric Hypothesis Testing (Theorem 7 and Corollary 8 of~\cite{cheng2024sample})] \label{thm:sample_C_no_private}
Fix the error probability $\alpha \in (0,pq)$. Then for non-orthogonal states $\rho$ and $\sigma$,
\begin{equation}
  \max\!\left\{ \frac{\ln(pq/\alpha) }{ -\ln F(\rho,\sigma) } ,\frac{1-\frac{\alpha(1-\alpha)}{pq}}{ \left[d_{\mathrm{B}}(\rho,\sigma)\right]^2  } \right\} \leq  \mathrm{SC}_{(\rho,\sigma)}(\alpha,p,q)  \leq \left \lceil \frac{ 2\ln \!\left( \frac{\sqrt{p q} }{ \alpha } \right) }{-\ln  F(\rho,\sigma)} \right\rceil.
\end{equation}
\end{theorem}

By using the fact $-\ln( \sqrt{x}) \geq 1- \sqrt{x}$ for all $x>0$, we have that $ -\ln  F(\rho,\sigma) \geq \left[d_B(\rho,\sigma)\right]^2$. Then, considering $\alpha,p,q$ to be constants and applying~\cref{thm:sample_C_no_private}, we arrive at 
\begin{equation} \label{eq:SC_no_privacy_d_B}
     \mathrm{SC}_{(\rho,\sigma)}(\alpha,p,q) = \Theta\! \left( \frac{1}{\left[d_B(\rho,\sigma)\right]^2}\right).
\end{equation}

\subsection{Private Quantum Hypothesis Testing} 

\label{Sec:private_hypothesis_testing}

In \cref{Sec:General_SC_o_privacy}, we reviewed the notion of sample complexity of symmetric quantum hypothesis testing and recalled bounds on it.
Now, we ask the question: how will this be affected if privacy of the quantum states is required? In particular, when we have access to the states $\cA(\rho)$ and $\cA(\sigma)$, where $\cA$ is a private quantum mechanism, how many samples of an unknown privatized state (i.e., $\cA(\rho)$ or $\cA(\sigma)$) do we need to achieve a fixed  error probability $\alpha$?

In this work, we employ QLDP as our privacy metric, which is defined in \cref{def:QLDP}.

Now, we are ready to formally define the sample complexity of private hypothesis testing. 
Let $\cA$ be an $\varepsilon$-QLDP mechanism. Then
\begin{align} \label{eq:def_SC}
    \mathrm{SC}^{\cA} _{(\rho,\sigma)}(\alpha,p,q) \coloneqq  
    & \min \!\left\{ n \in \mathbb{N}: p_{e}\!\left((\cA(\rho))^{\otimes n},(\cA(\sigma))^{\otimes n},p,q \right) \leq \alpha \right\},
\end{align}
where $p_e(\cdot)$ is defined in~\eqref{eq:probability_error}.
The sample complexity under $\varepsilon$-QLDP is defined as follows: 
\begin{equation}\label{eq:SC_private_int}
    \mathrm{SC}^{\varepsilon} _{(\rho,\sigma)} (\alpha,p,q)\coloneqq \inf_{\cA \in \cB^\varepsilon} \mathrm{SC}^{\cA} _{(\rho,\sigma)}(\alpha,p,q), 
\end{equation}
where $\cB^\varepsilon$ represents all $\varepsilon$-QLDP mechanisms, as defined in~\eqref{eq:epsilon_private_Channel}. 
We also use the shorthand $ \mathrm{SC}^{\varepsilon} _{(\rho,\sigma)}  $ to refer to the above quantity when $(\alpha,p,q)$ are clear from the context or when they are treated as constants.

\begin{remark}[Worst-Case Sample Complexity]
    When defining $ \mathrm{SC}^{\varepsilon} _{(\rho,\sigma)}$, we infimize over the set of channels,~$\cB^\varepsilon$. This corresponds to the best-case scenario. However, if we consider the worst-case scenario by instead taking the supremum over $\cB^\varepsilon$, the resulting sample complexity is unbounded. To see this, let us consider a replacement channel $\mathcal{R}$, where the channel replaces all inputs to the channel with a fixed, arbitrary state $\omega$. Then the channel described is indeed private with $\mathcal{R}(\rho)=\mathcal{R}(\sigma)=\omega$. However, in this construction, we cannot distinguish $\rho$ and $\sigma$ by having access to the privatized data (i.e., $\mathcal{R}(\rho)$ and $\mathcal{R}(\sigma)$), leading to an unbounded sample complexity. 
\end{remark}

Next, we provide upper and lower bounds on $\mathrm{SC}^{\varepsilon} _{(\rho,\sigma)} $.

\begin{theorem}[Bounds on Private Sample Complexity]\label{thm:bounds_sample_C_private}
    Let $p\in(0,1)$, set $q\coloneqq 1-p$, and let $\rho$ and $\sigma$ be states. Fix the error probability $\alpha \in (0,pq)$. For $\varepsilon>0$, the following holds:
    \begin{equation}
      \max \left\{\mathrm{SC}_{(\rho,\sigma)}(\alpha,p,q),\frac{C_{\varepsilon,p,q,\alpha}}{T(\rho,\sigma)}  \right \} 
      % \frac{C_{\varepsilon,p,q,\alpha}}{T(\rho,\sigma)} 
 \leq \mathrm{SC}^{\varepsilon} _{(\rho,\sigma)} (\alpha,p,q)\leq  \left \lceil 2 \ln\!\left( \frac{\sqrt{pq}}{\alpha} \right) \!\left( \frac{(e^\varepsilon +1)}{(e^\varepsilon -1) T(\rho, \sigma)} \right)^2  \right\rceil,
    \label{eq:private-sample-compl-bnds}
    \end{equation}
    where $\mathrm{SC}_{(\rho,\sigma)}(\alpha,p,q)$ is defined in~\eqref{eq:non_private_SC} and 
    \begin{equation}
        C_{\varepsilon,p,q,\alpha} \coloneqq   \max\!\left\{ \frac{\ln(pq/\alpha) (e^\varepsilon +1)}{\varepsilon (e^\varepsilon -1)  } ,\frac{\left({1-\frac{\alpha(1-\alpha)}{pq}} \right) (e^\varepsilon +1)}{2 \ (e^{\varepsilon/2}-1)^2  } \right\}.
    \end{equation}
\end{theorem}

\begin{IEEEproof}
First notice that, by applying a QLDP channel $\cA$, for $\varepsilon>0$ on states $\rho$ and $\sigma$, it is guaranteed that the states $\cA(\rho)$ and $\cA(\sigma)$ are not orthogonal. This is due to the fact that orthogonal states can be perfectly distinguished, which indeed violates the privacy requirement if it happens. With this, the current setup of privatized states satisfies the requirements needed to apply \cref{thm:sample_C_no_private}.

\underline{Upper bound:}
Notice that due to $-\ln(x) \geq 1- x$ for $x >0$
\begin{align}
    \frac{1}{-\ln \sqrt{F}(\rho,\sigma)} & \leq \frac{2}{\left[d_{\mathrm{B}}(\rho,\sigma)\right]^2 } 
    \leq \frac{2}{\left[T(\rho,\sigma)\right]^2 } ,
    \label{eq:neg-log-root-F-to-TD}
\end{align}
where the last inequality follows from the Fuchs-van-de-Graaf inequalities \cite{FG99}, as below:
\begin{equation}
    T(\rho,\sigma) \leq \sqrt{1- F(\rho,\sigma)} \leq \sqrt{2\left(1-\sqrt{F}(\rho,\sigma)\right)} = d_{\mathrm{B}}(\rho,\sigma).
\end{equation}

By fixing a private channel $\cA$, consider the distinguishability of the states $\cA(\rho)$ and $\cA(\sigma)$. To this end, by applying \cref{thm:sample_C_no_private} and~\eqref{eq:neg-log-root-F-to-TD}, we see that the sample complexity $\mathrm{SC} ^{\mathcal{A}}_{(\rho,\sigma)}$   satisfies the following:
\begin{equation}
    \mathrm{SC} ^{\mathcal{A}}_{(\rho,\sigma)} \leq \left \lceil   \frac{2\ln \!\left( \frac{\sqrt{p q} }{ \alpha } \right)}{\left[T(\cA(\rho), \cA(\sigma))\right]^2} \right \rceil,
\end{equation}
With that, we have 
\begin{equation}
    \mathrm{SC}^{\varepsilon} _{(\rho,\sigma)} \leq  \inf_{\cA \in \cB^\varepsilon}  \left \lceil    \frac{2 \ln \!\left( \frac{\sqrt{p q} }{ \alpha } \right)}{\left[T(\cA(\rho), \cA(\sigma))\right]^2} \right \rceil,
\end{equation}
This can further be bounded by choosing a specific channel that satisfies $\varepsilon$-QLDP. To this end choose $\cA=\cA^p_{\mathrm{Dep}} \circ \cM$ from \cref{prop: QLDP_mechanism} by setting $M= \Pi_{+}$, which is the projection onto the positive eigenspace of $\rho-\sigma$. 
% \mmw{double check notation}. 
This channel has also been used in the proof of \cref{thm:contraction_coeff_TD}. From there, apply~\eqref{eq:trace_dis_channelP} to get 
\begin{align}
 \mathrm{SC}^{\varepsilon} _{(\rho,\sigma)} 
   & \leq   \left \lceil  \frac{2 \ln \!\left( \frac{\sqrt{p q} }{ \alpha } \right)}{\left[T((\cA^p_{\mathrm{Dep}} \circ \cM)(\rho), (\cA^p_{\mathrm{Dep}} \circ \cM)(\sigma))\right]^2} \right \rceil\\
   & =  \left \lceil 2 \ln \!\left( \frac{\sqrt{p q} }{ \alpha } \right) \!\left( \frac{(e^\varepsilon +1)}{(e^\varepsilon -1) T(\rho, \sigma)} \right)^2 \right \rceil,
\end{align}
concluding the proof of the upper bound in~\eqref{eq:private-sample-compl-bnds}. 

\medskip 
\underline{Lower bound:}
To obtain the first part of the lower bound, observe that with the data processing of the channel $\cA^{\otimes n}$ and applying~\cite[Proposition~5.2]{khatri2020principles} we have 
\begin{equation}
   p_{e}\!\left(\rho^{\otimes n},\sigma^{\otimes n},p,q \right) \leq  p_{e}\!\left((\cA(\rho))^{\otimes n},(\cA(\sigma))^{\otimes n},p,q \right).
\end{equation}
Then using the definitions of sample complexity in~\eqref{eq:def_SC}, we have that 
\begin{equation}
     \mathrm{SC}_{(\rho,\sigma)}(\alpha,p,q) \leq   \mathrm{SC}_{(\rho,\sigma)}^\cA(\alpha,p,q).
\end{equation}
Next, optimizing over $\cA \in \cB^\varepsilon$, we arrive at 
\begin{equation}
  \mathrm{SC}_{(\rho,\sigma)}(\alpha,p,q)  \leq  \mathrm{SC}^{\varepsilon} _{(\rho,\sigma)} (\alpha,p,q).
\end{equation}

To obtain the second part of the lower bound in~\eqref{eq:private-sample-compl-bnds}, notice that 
\begin{equation}
    -\ln F(\rho,\sigma) = \widetilde{D}_{1/2}(\rho \Vert \sigma) \leq D(\rho \Vert \sigma),
\end{equation}
where $ \widetilde{D}_{1/2}$ is the sandwiched R\'enyi relative entropy of order $\alpha =1/2$~\cite{muller2013quantum, wilde2014strong}, and the last inequality follows from the $\alpha$-monotonicity of the sandwiched R\'enyi relative entropy, as well as the fact that $\lim_{\alpha \to 1} \widetilde{D}_{\alpha} = D$ (see also Proposition~7.28 and Proposition~7.29 in~\cite{khatri2020principles}). 
This leads to the following set of inequalities: 
\begin{align}
    \mathrm{SC}^{\varepsilon}_{ {(}
 \rho,\sigma)} &\geq \inf_{\cA \in \cB^\varepsilon} \frac{\ln(pq/\alpha)} { -\ln F\!\left(\cA(\rho),\cA(\sigma) \right) } \\ 
    & \geq \inf_{\cA \in \cB^\varepsilon} \frac{\ln(pq/\alpha)} { D\!\left(\cA(\rho) \Vert \cA(\sigma) \right) } \label{eq:SC_to_RE} \\ 
    & \geq  \ln\!\left( \frac{pq} {\alpha}\right) \frac{(e^\varepsilon +1)}{\varepsilon (e^\varepsilon -1) T(\rho,\sigma)},
\end{align}
where the last inequality follows from \cref{Cor:contraction_RE_TS}.

The second part of the lower bound follows by using \cref{prop:contraction_Bures_trace}  and the lower bound from \cref{thm:sample_C_no_private} as follows: 
\begin{align}
    \mathrm{SC}^{\varepsilon}_{{(}\rho,\sigma)} & \geq \inf_{\cA \in \cB^\varepsilon} \frac{pq-\alpha(1-\alpha)}{pq \left[d_{\mathrm{B}}(\cA(\rho),\cA(\sigma))\right]^2  } \\ 
   & \geq  \frac{\left(pq-\alpha(1-\alpha)\right) (e^\varepsilon +1)}{2 pq \ \left( e^{\varepsilon/2} -1 \right)^2 T(\rho,\sigma)  }. \label{eq:one_lower_bound}
\end{align}
Combining the two lower bounds completes the proof of the lower bound in~\eqref{eq:private-sample-compl-bnds}.
%%%%% We may need this 
\end{IEEEproof}

\begin{corollary}[Sample Complexity for Distinguishing Orthogonal States]\label{Cor:high_p_SC}
    Let $\rho$ and $\sigma$ be orthogonal quantum states. For $\varepsilon >0$ and $\alpha \leq pq$, by choosing $\alpha,p,q$ as constants,
    \begin{equation}\label{eq:SC_eps_all_ortho}
        \mathrm{SC}^{\varepsilon} _{(\rho,\sigma)} = \Theta\!\left(\left(\frac{e^\varepsilon +1}{e^\varepsilon -1}\right)^2 \right).
    \end{equation}
    Furthermore, in the high privacy regime (i.e., when $\varepsilon <1$), we obtain the following: 
    \begin{equation}
    \label{eq:SC_eps_all_ortho_high_priv}
        \mathrm{SC}^{\varepsilon} _{(\rho,\sigma)} = \Theta\!\left( \frac{1}{\varepsilon^2} \right).
    \end{equation}
\end{corollary}

\begin{IEEEproof}
First notice that
\begin{equation}
     \frac{(e^{\varepsilon/2}-1)^2}{e^\varepsilon+1}  \leq 
     \frac{(e^{\varepsilon/2}-1)^2}{e^\varepsilon +1} \times \frac{(e^{\varepsilon/2}+1 )^2}{e^{\varepsilon}+1}
     =\left(\frac{e^\varepsilon -1}{e^\varepsilon +1}\right)^2,
\end{equation}
with the first inequality following from $(e^{\varepsilon/2} +1)^2 \geq e^\varepsilon +1$. 
Then the lower bound in~\eqref{eq:one_lower_bound} simplifies to  
\begin{align}
    \mathrm{SC}^{\varepsilon}_{\rho,\sigma)}
   & \geq  \frac{\left(pq-\alpha(1-\alpha)\right) (e^\varepsilon +1)^2}{2 pq \ \left( e^{\varepsilon} -1 \right)^2 T(\rho,\sigma)  }.
\end{align}
Together with the upper bound in \cref{thm:bounds_sample_C_private} and the substitution $T(\rho, \sigma) =1$ for orthogonal states, we complete the first part of the proof.

    The equality in~\eqref{eq:SC_eps_all_ortho_high_priv} follows from applying~\eqref{eq:SC_eps_all_ortho}, along with the constraint $\varepsilon <1$, as follows: Firstly, we have
    \begin{equation}\label{eq:lower_highE}
         \frac{e^\varepsilon +1}{e^\varepsilon -1} %\geq \frac{e^\varepsilon}{e^\varepsilon -1}= \frac{1}{1-e^{-\varepsilon}}
         %=\frac{1}{\varepsilon -\varepsilon^2+\ldots} 
         \geq \frac{1}{\varepsilon},
    \end{equation}
    by observing that the left-hand side is equal to $\big(\tanh(\sfrac{x}{2})\big)^{-1}$ and applying $\tanh(\sfrac{x}{2}) \leq x$ for all $x>0$. Also 
    \begin{equation} \label{eq:upper_highE}
         \frac{e^\varepsilon +1}{e^\varepsilon -1} 
        \leq \frac{4}{\varepsilon}.
    \end{equation}  
    due to  $\tanh(\sfrac{x}{2}) \geq x/4$ for $x\in[0,1]$. This concludes the proof of~\eqref{eq:SC_eps_all_ortho_high_priv}.
\end{IEEEproof}

\begin{remark}[Order Optimality]
    In the setting discussed in \cref{Cor:high_p_SC} for orthogonal states, the $\varepsilon$-QLDP mechanism presented in \cref{prop: QLDP_mechanism} is order optimal (i.e., optimal up to constant factors) for all $\varepsilon>0$.
\end{remark}

\begin{remark}[Cost of Privacy] \label{rem:cost_privacy_high}
    For orthogonal states, we need only one sample of the unknown state to declare whether it is $\rho$ or $\sigma$ when we have access to non-privatized samples of $\rho$ or $\sigma$. However, to achieve $\varepsilon$-QLDP, the input states are privatized by applying the channel $\cA$, so that we have access to  $\cA(\rho)$ and $\cA(\sigma)$ only. To this end, we need samples on the order of $\Theta\! \left(\left((e^\varepsilon +1)/(e^\varepsilon-1)\right)^2\right)$ for all $\varepsilon > 0$, which is apparent from \cref{Cor:high_p_SC}.

    In addition, by \cite[Proposition~10]{nuradha2023quantum}, for an $\varepsilon$-QLDP mechanism $\cA$ and  for all  states $\rho$ and $\sigma$, we have 
    \begin{equation}
        D\!\left( \cA(\rho) \Vert \cA(\sigma) \right) \leq \min\!\left\{ \varepsilon, \frac{\varepsilon^2}{2} \right\}.
    \end{equation}
Applying that in the proof of the lower bound of \cref{thm:bounds_sample_C_private} (in particular, in~\eqref{eq:SC_to_RE}), it follows that, for all distinct states $\rho$ and $\sigma$,  we require 
\begin{equation}
    {\mathrm{SC}}^{\varepsilon} _{(\rho,\sigma)} = \Omega\!\left( \frac{1}{\min\!\left\{ \varepsilon, {\varepsilon^2} \right\}}\right),
\end{equation}
in order to ensure $\varepsilon$-QLDP, where $\varepsilon>0$.
\end{remark}

\cref{rem:cost_privacy_high} discusses the cost of privacy when we have access to privatized data samples instead of original data. Next, we show that when certain conditions are met, the impact of privacy on the sample complexity is not significant and is comparable to the non-private sample complexity. {This also provides a sense of how large  the privacy parameter $\varepsilon$ should be in order to achieve a similar sample complexity as in the non-private case, where we do not need to pay an extra cost to ensure privacy, $d_B(\rho,\sigma) > 1$.}

\begin{proposition}[Sample Complexity in the Low-Privacy Regime]\label{prop:SC_low_privacy}
    Let $\rho$ and $\sigma$ be qubit states, and let $\varepsilon >0$. If
\begin{equation}
    \left( \frac{e^\varepsilon -1}{e^\varepsilon +1}\right)^2 \geq \frac{1}{\left[d_B(\rho,\sigma)\right]^2}, 
\end{equation}
then
\begin{equation}
     {\mathrm{SC}}^{\varepsilon} _{(\rho,\sigma)}=\Theta\!\left(  \frac{1}{ \left[d_B(\rho,\sigma)\right]^2} \right).
\end{equation}
Furthermore, for $\rho$ and $\sigma$ general quantum states, 
if
\begin{equation}
    \left( \frac{e^\varepsilon -1}{e^\varepsilon +1}\right)^2 \geq \frac{1}{\left[d_B(\rho,\sigma)\right]^2}, 
\end{equation}
then
\begin{equation}
    \Omega\!\left( \frac{1}{ \left[d_B(\rho,\sigma)\right]^2}\right)\leq    {\mathrm{SC}}^{\varepsilon} _{(\rho,\sigma)} \leq O \!\left(\frac{ L_{k,k'}}{\left[d_B(\rho,\sigma)\right]^2} \right),
\end{equation}
where
\begin{equation}\label{eq:L_k_k_'}
    L_{k,k'} \coloneqq \left(\max\!\left\{1, \frac{\min\{k,k'\}}{2} \right\}\right)^2,
\end{equation}
$k' \coloneqq \ln\!\left(4/\left[d_B(\rho,\sigma)\right]^2\right)$, and $k$ is the number of 
distinct eigenvalues of the operator $\rho \# \sigma^{-1}$, as defined in~\cref{eq:geometric-mean}.
\end{proposition}
\begin{IEEEproof}
    See~\cref{app:SC_low}.
\end{IEEEproof}

\medskip
Note that from~\cref{prop:SC_low_privacy}, when $\rho$ and $\sigma$ are qubits, the order optimality in sample complexity is achieved by the privatization mechanism presented in~\cref{prop: QLDP_mechanism} by choosing the measurement channel to correspond to the measurement in the eigenbasis of $\rho \# \sigma^{-1}$. 

\begin{remark}[Different Private Mechanisms] \label{rem:different_private_mechanisms}
 Previously, we considered the setting in which the same private mechanism $\cA \in \cB^\varepsilon$ is applied to each sample of $\rho$ or $\sigma$. We can also define another variant in which different mechanisms are applied at each stage; namely, $\cA_i \in \cB^\varepsilon$ for all $i \in \{1,\ldots,n\}$. Then, we define the associated sample complexity as follows:
    \begin{equation}\widetilde{\mathrm{SC}}^{\varepsilon} _{(\rho,\sigma)} \coloneqq \inf_{\substack{\cA_i \in \cB^\varepsilon \\  \forall i \in \{1,\ldots,n\}} }\min\! \left\{ n \in \mathbb{N} : p_{e}\!\left( \cA_1(\rho) \otimes \cdots \otimes \cA_n(\rho) ,\cA_1(\sigma) \otimes \cdots \otimes \cA_n(\sigma),p,q \right) \leq  \alpha \right\}, 
\end{equation}
By choosing each $\cA_i$ to be the same mechanism, we conclude the following inequality: 
\begin{equation} \widetilde{\mathrm{SC}}^{\varepsilon} _{(\rho,\sigma)} \leq {\mathrm{SC}}^{\varepsilon} _{(\rho,\sigma)}.
\end{equation}
With that, the upper bound given in \cref{thm:bounds_sample_C_private} still holds. 
However, the lower bound may not hold. To this end, we prove the following: for $\alpha \leq \min\{p,q\}$
\begin{equation}\label{eq:lower_bound_different_mechanism}
\widetilde{\mathrm{SC}}^{\varepsilon} _{(\rho,\sigma)} \geq 2 \left( 1- \frac{\alpha}{\min\{p,q\}}\right)^2 \frac{(e^\varepsilon + 1)}{\varepsilon(e^\varepsilon -1) T(\rho, \sigma)}.
\end{equation}
The proof is presented in Appendix~\ref{APP:proof_different_mechanisms}.\footnote{Note that this lower bound also holds in the case where we consider the same private channel for every copy of the state.}

Under this setting, also for orthogonal states $\rho$ and $\sigma$ and choosing $\alpha,p,q$ as constants, when $\varepsilon<1$, we arrive at 
\begin{equation}
\widetilde{\mathrm{SC}}^{\varepsilon} _{(\rho,\sigma)} = \Theta\!\left(\frac{1}{\varepsilon^2} \right),
\end{equation}
implying that having different mechanisms may not improve the sample complexity for this case.

Above, we considered $\cA_i$ chosen independently of $\cA_j$, where $i\neq j$. 
It is an interesting question for future work to understand whether adaptive strategies can decrease the sample complexity or number of copies of the privatized states required to minimize the cost of privacy.
\end{remark}

\begin{remark}[Private Multiple Hypothesis Testing]\label{rem:multiple_hypo_private}
    The sample complexity of private multiple hypothesis testing for a tuple of states $(\rho_m)_{m=1}^M$ with the prior probabilities $(p_m)_{m=1}^M$ to achieve at most $\alpha$ error probability is defined as 
    \begin{equation}
       {\mathrm{SC}}^{\varepsilon} _{(\rho_i)_{i=1}^M} \coloneqq \inf_{\cA \in \cB^\varepsilon} \left\{ n \in \mathbb{N}:  \inf_{ (\Lambda^{(n)}_1, \ldots, \Lambda^{(n)}_M) } \sum_{m=1}^M p_m \Tr\!\left[(I^{\otimes n}-\Lambda^{(n)}_m) \left(\cA(\rho_m)\right)^{\otimes n} \right] \leq \alpha \right\},
    \end{equation}
where $\Lambda^{(n)}_1,\ldots, \Lambda^{(n)}_M \geq 0$ and $\sum_{m=1}^M \Lambda^{(n)}_m = I^{\otimes n}$.
Then, with the use of tools from the proof of~\cref{thm:bounds_sample_C_private} and~\cite[Theorem~11]{cheng2024sample}, we arrive at 
\begin{equation}
    	\max_{m\neq \bar{m}}  \frac{ \ln\!\left( \frac{p_m p_{\bar m}}{ (p_m + p_{\bar m})\alpha } \right) (e^\varepsilon +1)}{ \varepsilon (e^\varepsilon -1) T(\rho_m,\rho_{\bar{m}})}
		\leq  {\mathrm{SC}}^{\varepsilon} _{(\rho_i)_{i=1}^M} 
	 \leq 
		\left\lceil  \max_{m\neq \bar{m}}  {2\ln\!\left( \frac{ M(M-1) \sqrt{p_m} \sqrt{p_{\bar{m}}}  }{ 2\alpha } \right) }\left( \frac{(e^\varepsilon +1)}{(e^\varepsilon -1) T(\rho_m, \rho_{\bar{m}})} \right)^2\right\rceil .
\end{equation}
\end{remark}

\begin{remark}[Private Asymmetric Hypothesis Testing]\label{rem:asym_hypoth_SC_lower}
We define the private asymmetric hypothesis testing sample complexity as follows: Let $\rho$ and $\sigma$ be states with $\alpha_1, \alpha_2 \in (0,1)$ and $\varepsilon \geq 0$
\begin{align}
n^{\ast}(\rho,\sigma,\varepsilon,\alpha_1, \alpha_2)  & = \inf_{\cA \in \cB^{\varepsilon}}\inf_{\Lambda^{(n)}}\left\{
\begin{array}
[c]{c}
n\in\mathbb{N}:\operatorname{Tr}[(I^{\otimes n}-\Lambda^{(n)}) \cA(\rho)^{\otimes
n}]\leq\alpha_1,\\
\operatorname{Tr}[\Lambda^{(n)} \cA(\sigma)^{\otimes n}]\leq\alpha_2,\ 0\leq
\Lambda^{(n)}\leq I^{\otimes n}
\end{array}
\right\},
\label{eq:asymm-beta-rewrite-both-errs}
\end{align}
where $\cB^\varepsilon$ is defined in~\eqref{eq:epsilon_private_Channel}.
A lower bound on $n^{\ast}(\rho,\sigma,\varepsilon,\alpha_1, \alpha_2) $ is as follows:
    \begin{equation}
        n^{\ast}(\rho,\sigma,\varepsilon,\alpha_1, \alpha_2)  \geq \max\!\left\{  \sup_{ \beta >1 } \left(  \frac{\ln\!\left(  \frac{\left(  1-\alpha_1 \right)
^{\beta^{\prime}}}{\alpha_2}\right)  }{\min\{ \varepsilon, \varepsilon^2 \beta /2\}}\right)  ,\ \sup_{ \beta >1 } \left(  \frac{\ln\!\left(  \frac{\left(
1-\alpha_2\right)  ^{\beta^{\prime}}}{\alpha_1}\right)  }{\min\{ \varepsilon, \varepsilon^2 \beta /2\}}\right)
%, n^{\ast}(\rho,\sigma,\infty,\alpha_1, \alpha_2) 
\right\},
\label{eq:binary_asymmetric_samp_comp}
    \end{equation}
where $\beta' \coloneqq \frac{\beta}{\beta-1}$. 
    The proof follows by applying~\cite[Proposition~9]{nuradha2023quantum} together with the lower bound presented in~\cite[Theorem~9]{cheng2024sample}. In particular, we use the following inequality: Let $\cA \in \cB^\varepsilon$ and $\beta >1$
    \begin{equation}
        \widetilde
{D}_{\beta}\!\left(\cA(\rho)\Vert\cA(\sigma)\right) \leq \min\! \left\{ \varepsilon, \frac{\varepsilon^2 \beta}{2}\right\} < + \infty,
    \end{equation}
where $ \widetilde
{D}_{\beta}(\omega \Vert \tau) \coloneqq \frac{1}{\beta - 1} \ln \operatorname{Tr}[(\tau^{(1-\beta)/2\beta} \omega \tau^{(1-\beta)/2\beta})^{\beta}]$ is the sandwiched R\'enyi relative entropy of order~$\beta$~\cite{muller2013quantum, wilde2014strong}. 
\end{remark}

{
\begin{remark}[Comparison to~\cite{Christoph2024sample}]
    The key difference of our findings and the independent work [36] is that our sample complexity bounds in~\cref{thm:bounds_sample_C_private} hold for arbitrary $\alpha, p,q$ such that $\alpha \in (0,pq)$, while their results are tailored for specific values (see Theorem~4.10 and Eq.~(1.6) and Eq.~(1.7) in \cite{Christoph2024sample}). If we also choose those specific values that they consider, our upper bound on the sample complexity coincides with theirs. For the lower bound, our bound is tighter when considering the dependence on the privacy parameter $\varepsilon$ while their bound is tighter when considering the dependence on
    $T(\rho,\sigma)$.  Our lower bound is a key tool in precisely characterizing the sample complexity of orthogonal quantum states when privatized quantum states are given to us, as we have concluded in~\cref{Cor:high_p_SC}.
\end{remark}
}
 
\subsection{Instance-Specific Private Channels and Sample Complexity} \label{Sec:instance_Specific_eps}

In the discussions above, we considered the set of all $\varepsilon$-QLDP mechanisms, denoted by $\cB^\varepsilon$, and chose the best mechanism from this set that requires the smallest number of samples for the hypothesis testing task. Next, we select an instance-specific set of channels, denoted by $\mathcal{W}^{\varepsilon}_{(\rho,\sigma)}$ (defined in~\eqref{eq:special_set_channels}) and characterize the sample complexity in distinguishing two states $\rho$ and $\sigma$. For $\varepsilon \geq 0$,
define 
\begin{equation}\label{eq:special_set_channels}
\mathcal{W}^{\varepsilon}_{(\rho,\sigma)} \coloneqq  
\left \{ \cA \in \cB^\varepsilon: \begin{aligned} &  \max\{ \lambda_{\min}\!\left( \cA(\rho)\right), \lambda_{\min}\!\left( \cA(\sigma)\right)\} \geq \frac{1}{e^\varepsilon +1}
\end{aligned} \right \} ,
\end{equation}
where $\lambda_{\min}(\omega)$ denotes the minimum eigenvalue of $\omega$.

Although a minimum eigenvalue constraint might seem restrictive, we note here that this constraint is imposed on the \textit{output states} after applying private mechanisms, where the private channel may map the input state to a lower dimensional space.
Furthermore, this class of channels is non-empty, and it also includes practically relevant channels. To this end, the depolarization mechanism with flip parameter $p>0$ belongs to this class for all states $\rho$ and $\sigma$ under certain conditions. Moreover, the mechanism proposed in \cref{prop: QLDP_mechanism} also belongs to this class, and we prove it formally in  the proof of \cref{thm:instance_sample_C_private} below.

Let us now define the private sample complexity achieved by ensuring privacy through the set of channels $\mathcal{W}^{\varepsilon}_{(\rho,\sigma)}$: 
\begin{equation} \widehat{\mathrm{SC}}^{\varepsilon} _{(\rho,\sigma)} \coloneqq \inf_{\cA \in \mathcal{W}^{\varepsilon}_{(\rho,\sigma)}} \mathrm{SC}^{\cA} _{(\rho,\sigma)}(\alpha,p,q).
\end{equation}
By definition, the following inequality holds (whenever $\mathcal{W}^{\varepsilon}_{(\rho,\sigma)}$ is non-empty):%, which is true for distinct pairs of states):
\begin{equation}
{\mathrm{SC}}^{\varepsilon} _{(\rho,\sigma)} \leq \widehat{\mathrm{SC}}^{\varepsilon} _{(\rho,\sigma)}.
\end{equation}

\begin{proposition}
[Instance-Specific Private Sample Complexity Bounds] \label{thm:instance_sample_C_private}
 Let $p\in(0,1)$ and set $q\coloneqq 1-p$. Fix the error probability $\alpha \in (0,pq)$.
 Let $\rho$ and $\sigma$ be distinct quantum states. 
   The  following inequalities hold
    \begin{equation}
    \label{eq:instance-specific-SC-bnds}
     \ln\!\left( \frac{pq} {\alpha}\right) \frac{1}{e^\varepsilon+1}\left(\frac{(e^\varepsilon +1)}{ (e^\varepsilon -1) T(\rho,\sigma)} \right)^2 \leq \widehat{\mathrm{SC}}^{\varepsilon} _{(\rho,\sigma)} \\ 
    \leq \ln\!\left( \frac{\sqrt{pq}}{\alpha} \right) 
    \left(\frac{(e^\varepsilon +1)}{(e^\varepsilon -1) T(\rho, \sigma)}\right)^2.
    \end{equation}
Furthermore, for the high privacy regime $\varepsilon <1$ and choosing $ p,q$ to be constants, we conclude that
     \begin{equation}  \widehat{\mathrm{SC}}^{\varepsilon} _{(\rho,\sigma)} =\Theta \! \left( \left(\frac{1}{\varepsilon T(\rho,\sigma)}\right)^2  \right).
     \end{equation} 
\end{proposition}

\begin{IEEEproof}
See Appendix~\ref{App:Proof_instance_SC}.
\end{IEEEproof}

\begin{remark}[Impact of Dimension with a Less Restricted Setting]
By choosing 
\begin{equation}
\widehat{\mathcal{W}}^{\varepsilon}_{(\rho,\sigma)} \coloneqq  
\left \{ \cA \in \cB^\varepsilon: \begin{aligned} &  \max\{ \lambda_{\min}\!\left( \cA(\rho)\right), \lambda_{\min}\!\left( \cA(\sigma)\right)\} \geq \frac{1}{e^\varepsilon +d-1}
\end{aligned} \right \} , 
\end{equation}
we arrive at the following upper and lower bounds on the sample complexity
\begin{equation}\label{eq:SC_with_d_con}
    \ln\!\left( \frac{pq} {\alpha}\right) \frac{1}{e^\varepsilon+d-1}\left(\frac{(e^\varepsilon-1)}{ (e^\varepsilon -1) T(\rho,\sigma)} \right)^2 \leq \inf_{\cA \in \widehat{\mathcal{W}}^{\varepsilon}_{(\rho,\sigma)}} \mathrm{SC}^{\cA} _{(\rho,\sigma)}(\alpha,p,q) \\ 
    \leq \ln\!\left( \frac{\sqrt{pq}}{\alpha} \right) 
    \left(\frac{(e^\varepsilon +d-1)}{(e^\varepsilon -1) T(\rho, \sigma)}\right)^2.
\end{equation}
The lower bound follows similarly to the proof of~\cref{thm:instance_sample_C_private}. To obtain the upper bound choose the following $\varepsilon$-QLDP mechanism: $\cA^p_{\mathrm{Dep}}$ with $p \geq d/(e^\varepsilon +d-1)$ satisfies $\varepsilon$-QLDP (\cite[Eq.~(129)]{nuradha2023quantum}). The class of channels in $\widehat{\mathcal{W}}^{\varepsilon}_{(\rho,\sigma)} $ is a larger set compared to~\eqref{eq:special_set_channels}. However, the gap between the upper and lower bounds in~\eqref{eq:SC_with_d_con} increases with $d$, the dimension of the states.
\end{remark}

\begin{remark}[Order Optimality]
    In the setting discussed in \cref{thm:instance_sample_C_private} with $\varepsilon<1$, the $\varepsilon$-QLDP mechanism presented in \cref{prop: QLDP_mechanism} achieves order-optimality (optimal up to constant factors) in terms of sample complexity. 
\end{remark}

{In summary, this section shows that there exist sets of private channels that depend on the states chosen (defined in~\eqref{eq:special_set_channels}), where the general sample complexity bounds stated in~\cref{thm:bounds_sample_C_private} can be improved as in~\cref{thm:instance_sample_C_private} to minimize the gap between the lower and upper bounds. We leave further analysis on practically relevant and operationally
motivated sets of channels achieving similar improvements for future work.}

\section{Contraction under $(\varepsilon,\delta)$-QLDP channels}

\label{Sec:Contraction_eps_delta}

In the previous sections, we focused on privacy constraints imposed by $\varepsilon$-QLDP with $\delta=0$. Generalizing~\eqref{eq:contraction_generalized_divergence}, set $\delta \geq 0$, and define privatized contraction coefficients under $(\varepsilon, \delta)$-QLDP privacy constraints  as follows: 
%similar to~: 
\begin{equation}
    \eta_{\boldsymbol{D}}^{\varepsilon,\delta} \coloneqq \sup_{\substack{\cN \in \cB^{\varepsilon, \delta}, \\ \rho, \sigma \in \cD, \\ \boldsymbol{D}(\rho \Vert \sigma) \neq 0 }} \frac{\boldsymbol{D} \!\left( \cN(\rho) \Vert \cN(\sigma)\right)}{\boldsymbol{D}(\rho \Vert \sigma)},
\end{equation}
where $\mathcal{B}^{\varepsilon,\delta}$ corresponds to the set of all $(\varepsilon,\delta)$-QLDP mechanisms, defined formally as
\begin{equation} \label{eq:epsilon_delta_private_Channel}
\mathcal{B}^{\varepsilon,\delta} \coloneqq \left\{ \cN \in \operatorname{CPTP}: \sup_{\rho,\sigma \in \cD(\cH)} E_{e^\varepsilon}\!\left( \cN(\rho) \Vert \cN(\sigma) \right) \leq \delta \right\} . 
\end{equation} 
Next, we quantify the contraction of the normalized trace distance by instantiating the generalized divergence to be the normalized trace distance. The following extends~\cref{thm:contraction_coeff_TD} and provides an alternative proof for the converse bound in the proof of~\cref{thm:contraction_coeff_TD}.

\begin{theorem}[Contraction of Trace Distance under  $(\varepsilon,\delta)$-QLDP Constraints]\label{thm:contraction_T_eps_delta}
Let $\varepsilon,\delta \geq 0$, and let $\rho$ and $\sigma$ be states such that $T(\rho,\sigma) \neq 0$. Then
 \begin{equation}
        \sup_{\cN \in \cB^{\varepsilon,\delta}} \frac{T \!\left( \cN(\rho), \cN(\sigma)\right)}{T (\rho, \sigma)} = \frac{e^\varepsilon -1 + 2\delta}{e^\varepsilon +1}.
    \end{equation}
   Consequently, the privatized contraction coefficient for the trace distance under $(\varepsilon,\delta)$-QLDP is given by 
    \begin{equation}
        \eta^{\varepsilon,\delta}_{T} = \frac{e^\varepsilon -1+2 \delta}{e^\varepsilon +1 }.
    \end{equation}
\end{theorem}

\begin{IEEEproof}
\underline{Converse bound:}
    Let $\rho$ and $\sigma$ be states, let $\gamma \geq 1$, and consider that 
    \begin{align}
        (\gamma+1) T(\rho,\sigma) &=  (\gamma+1) \sup_{0 \leq M \leq I} \Tr\! \left[ M (\rho- \sigma)\right]\\ 
        &= \sup_{0 \leq M \leq I} \left\{\Tr[M(\rho-\gamma \sigma)] + \Tr[M(\gamma\rho- \sigma)]\right\} \\ 
        &\leq \sup_{0 \leq M \leq I} \left\{\Tr[M(\rho-\gamma \sigma)]\right\} + \sup_{0 \leq M \leq I} \left\{\Tr[M(\gamma\rho- \sigma)] \right\} \\ 
         &= E_\gamma(\rho \Vert \sigma) + \gamma \sup_{0 \leq M \leq I} \Tr\!\left[M\left(\rho- \frac{1}{\gamma}\sigma\right)\right] \\ 
        % & = E_\gamma(\rho \Vert \sigma) + \gamma E_{\frac{1}{\gamma}}(\rho \Vert \sigma) \\
        &= E_\gamma(\rho \Vert\sigma) +E_\gamma(\sigma \Vert \rho) + \gamma -1,
    \end{align}
where the penultimate equality follows from the variational representation of the hockey-stick divergence from~\cite[Lemma~II.1]{hirche2023quantum} and the last equality follows from the following reasoning (see also the symmetry property of hockey-stick divergence in \cite[Eq.~(II.21)]{hirche2023quantum}):
\begin{align}
  E_\gamma(\sigma \Vert \rho) &=  \sup_{0 \leq M \leq I} \Tr[M(\sigma-\gamma \rho)] \\ 
  &= \sup_{0 \leq M \leq I} \Tr[(I-M) (\sigma-\gamma \rho)]\\
  &= \Tr[\sigma-\gamma \rho] + \sup_{0 \leq M \leq I} \Tr[M(\gamma \rho-\sigma)] \\
  &=(1-\gamma)+ \gamma \sup_{0 \leq M \leq I} \Tr\!\left[M\left(\rho- \frac{1}{\gamma}\sigma\right)\right].
\end{align}

With the above setup, choose $\rho =\cA(|\phi\rangle\!\langle \phi|)$ and $\sigma= \cA(|\psi\rangle\!\langle \psi|)$, where $\cA \in \cB^{\varepsilon,\delta}$ and $|\phi\rangle\!\langle \phi|, |\psi\rangle\!\langle \psi|$ are orthogonal pure states together with $\gamma= e^\varepsilon$. Then, we have 
\begin{align}
   & (e^\varepsilon +1) T\!\left( \cA(|\phi\rangle\!\langle \phi|), \cA(|\psi\rangle\!\langle \psi|) \right) \notag  \\
   & \leq E_{e^\varepsilon}\!\left( \cA(|\phi\rangle\!\langle \phi|) \Vert  \cA(|\psi\rangle\!\langle \psi|) \right) + E_{e^\varepsilon}\!\left( \cA(|\psi\rangle\!\langle \psi|) \Vert  \cA(|\phi\rangle\!\langle \phi|) \right) + e^\varepsilon -1 \\ 
    &\leq 2 \delta + e^\varepsilon -1,
\end{align}
where the last inequality follows due to $\cA \in \cB^{\varepsilon,\delta}$ and applying the constraints $E_{e^\varepsilon}(\rho \Vert \sigma) \leq \delta$ and $E_{e^\varepsilon}(\sigma \Vert \rho) \leq \delta$. We arrive at 
\begin{equation}\label{eq:trace_eps_delta}
    T\!\left( \cA(|\phi\rangle\!\langle \phi|), \cA(|\psi\rangle\!\langle \psi|) \right) \leq \frac{2 \delta + e^\varepsilon -1}{e^\varepsilon +1}.
\end{equation}

Next, with the choice $\gamma=1$ in~\eqref{eq:hockey-stick-contraction-general}, we have for all $\cA \in \cB^{\varepsilon, \delta}$
\begin{align}
     \sup_{\substack{\rho, \sigma \in \cD }} \frac{T\!\left( \cA(\rho), \cA(\sigma)\right)}{T(\rho, \sigma)} 
  &= \sup_{| \phi \rangle \perp  | \psi \rangle}  T\!\left( \cA(| \phi\rangle\!\langle \phi|) , \cA(| \psi \rangle\!\langle \psi|) \right) \\
  & \leq \frac{2 \delta + e^\varepsilon -1}{e^\varepsilon +1},
\end{align}
where the last inequality follows from~\eqref{eq:trace_eps_delta}.

So we conclude that 
\begin{equation}\label{eq:upper_bound_eps_delta}
     \sup_{\cA \in \cB^{\varepsilon,\delta}} \frac{T \!\left( \cA(\rho), \cA(\sigma)\right)}{T (\rho, \sigma)}  \leq \eta_T^{\varepsilon,\delta} \leq \frac{2 \delta + e^\varepsilon -1}{e^\varepsilon +1}.
\end{equation}

\medskip 
\underline{Achievability bound:}
Let $\cA= \cA^p_{\mathrm{Dep}} \circ \cM$ with $p=2(1-\delta)/ (e^\varepsilon +1)$, where $\cM$ corresponds to~\eqref{eq:measurement_channel}. First, we need to show that $\cA \in \cB^{\varepsilon,\delta}$. To this end, consider that
\begin{align}
    E_{e^\varepsilon}\left(\cA(\rho) \Vert \cA(\sigma) \right)& \leq \max\!\left \{0, (1-e^\varepsilon) \frac{p}{2} +(1-p) E_{e^\varepsilon} \left(\cM(\rho) \Vert \cM(\sigma)\right) \right\} \\ 
    &\leq \max\!\left \{0, (1-e^\varepsilon) \frac{p}{2} +(1-p)  \right\} \\
    & \leq \max\{0,\delta\} \\
    &=\delta,
\end{align}
where the first inequality follows from~\cite[Lemma~IV.1]{hirche2023quantum}, the second inequality from $ E_{e^\varepsilon} \left(\cM(\rho) \Vert \cM(\sigma)\right) \leq 1$, and the last inequality by substituting $p=2(1-\delta)/(e^\varepsilon +1)$. This shows that for all $\rho,\sigma$, $  E_{e^\varepsilon}\left(\cA(\rho) \Vert \cA(\sigma) \right) \leq \delta$, proving that $\cA \in  \cB^{\varepsilon,\delta}$.

Now choose $M$ in~\eqref{eq:measurement_channel} to be the projection onto the positive eigenspace of $\rho-\sigma$. With that we have 
\begin{align}
     T\!\left( \cA(\rho), \cA(\sigma)\right) &=(1-p) T(\rho,\sigma) \\ 
     &= \frac{2 \delta + e^\varepsilon -1}{e^\varepsilon +1} T(\rho,\sigma),
\end{align}
showing that for all $\rho,\sigma$ such that $T(\rho,\sigma) \neq 0$ with the specific choice of $\cA= \cA^p_{\mathrm{Dep}} \circ \cM$, we achieve the upper bound in~\eqref{eq:upper_bound_eps_delta}, concluding the proof.
\end{IEEEproof}

\begin{corollary}[Contraction of $f$-divergences under $(\varepsilon,\delta)$-QLDP]
Let $\varepsilon, \delta \geq 0$ with $\cA$ satisfying $(\varepsilon,\delta)$-QLDP. For states $\rho$ and $\sigma$, we have 
\begin{equation}
    D_f\!\left( \cA(\rho) \Vert \cA(\sigma) \right) \leq \frac{e^\varepsilon -1 +2 \delta}{e^\varepsilon +1} D_f(\rho \Vert \sigma),
\end{equation}
where $D_f(\cdot\Vert \cdot)$ is defined in~\eqref{eq:f_divergence}.   
\end{corollary}
\begin{IEEEproof}
    Proof follows by applying the fact that the contraction coefficient of the $f$-divergence is upper bounded by the contraction coefficient of trace distance~\cite[Lemma~4.1]{hirche2023quantum_2} together with~\cref{thm:contraction_T_eps_delta}.
\end{IEEEproof}
\begin{remark}[Improvement Compared to Existing Results]
  \cite[Corollary~V.I]{hirche2023quantum} states the following upper bound:
  \begin{equation}\label{eq:eps_delta_T_previous}
       \eta^{\varepsilon,\delta}_{T}  \leq \frac{e^\varepsilon -1+ \delta}{e^\varepsilon }. 
  \end{equation}
\cref{thm:contraction_T_eps_delta} provides a tight characterization of the privatized contraction coefficient under $(\varepsilon, \delta)$-QLDP constraints, which can be seen from 
\begin{equation}
    \frac{e^\varepsilon -1+ \delta}{e^\varepsilon } - \frac{e^\varepsilon -1+2 \delta}{e^\varepsilon +1 } = \frac{(1-\delta) (e^\varepsilon -1)}{e^\varepsilon (e^\varepsilon +1)} \geq 0.
\end{equation}
Note that \cref{thm:contraction_T_eps_delta} also improves the best known bound for the contraction of total variation distance in the classical setting (See \cite[Lemma~1]{asoodeh2021local}). 

By using~\cref{thm:contraction_T_eps_delta} instead of~\eqref{eq:eps_delta_T_previous}, it is possible to improve the upper bound on the 
error exponent in asymmetric hypothesis testing with privatized quantum states ($\cA(\rho)$  and $\cA(\sigma)$), as presented in~\cite[Corollary~5.14]{hirche2023quantum_2}, as follows: For all $\cA$ satisfying $(\varepsilon, \delta)$-QLDP we have that
\begin{equation}
    \lim_{n \to \infty} -\frac{1}{n} \ln \beta_{\nu}\!\left( \cA(\rho)^{\otimes n} \Vert \cA(\sigma)^{\otimes n} \right)= D\!\left(\cA(\rho)\Vert \cA(\sigma) \right) \leq  \frac{e^\varepsilon -1+2 \delta}{e^\varepsilon +1 } D(\rho \Vert \sigma),
\end{equation}
where $\nu \in (0,1)$ and 
\begin{equation} 
\beta_{\nu}\!\left(\cA(\rho)^{\otimes n}\Vert\cA(\sigma)^{\otimes n}\right)\coloneqq\inf
_{\Lambda^{(n)}}\left\{
\begin{array}
[c]{c}
\operatorname{Tr}[\Lambda^{(n)}\cA(\sigma)^{\otimes n}]:\\\operatorname{Tr}
[(I^{\otimes n}-\Lambda^{(n)})\cA(\rho)^{\otimes n}]\leq\nu,\\
0\leq\Lambda^{(n)}\leq I^{\otimes n}
\end{array}
\right\}  .
\label{eq:beta-err-asymm}
\end{equation}
\end{remark}

\begin{remark}[Private Quantum Hypothesis Testing under $(\varepsilon,\delta)$-QLDP] 
In \cref{thm:bounds_sample_C_private}, we provide bounds on the sample complexity of private quantum hypothesis testing where privacy constraints consist of $\varepsilon$-QLDP with $\delta=0$. For the case where $\delta \geq 0$, the upper bound therein can be generalized as follows:
\begin{equation}
    \inf_{\cA \in \cB^{\varepsilon,\delta}} \mathrm{SC}^{\cA} _{(\rho,\sigma)}(\alpha,p,q) \leq \left \lceil 2 \ln\!\left( \frac{\sqrt{pq}}{\alpha} \right) \!\left( \frac{(e^\varepsilon +1)}{(e^\varepsilon -1+2\delta) T(\rho, \sigma)} \right)^2  \right\rceil.
\end{equation}
This follows by the use of the mechanism introduced in the achievability part of the proof of~\cref{thm:contraction_T_eps_delta} together with similar techniques used in the proof of the upper bound in~\cref{thm:bounds_sample_C_private}. {For the setting of instance specific private channels in~\cref{Sec:instance_Specific_eps}, it is possible to extend \cref{thm:instance_sample_C_private} to the $\delta \neq 0$ setting with the use of the contraction of trace distance in~\cref{thm:contraction_T_eps_delta}.}

We leave the general characterization of sample complexity of hypothesis testing when privacy is imposed by $(\varepsilon,\delta)$-QLDP for future work. {The main technical tools needed to obtain the lower bounds as given in~\cref{thm:bounds_sample_C_private} are the characterization of the contraction of quantum relative entropy and Bures distance under the privacy constraints imposed by $(\varepsilon,\delta)$-QLDP.}
In~\cref{app:other_connections_eps_delta}, we provide some tools that may be of relevance to accomplish this goal. In particular, we prove several connections between $\varepsilon$-QLDP and $(\varepsilon,\delta)$-QLDP mechanisms, which may potentially be used to infer about the impact of $(\varepsilon,\delta)$-QLDP constraints using the formal guarantees derived for $\varepsilon$-QLDP in~\cref{thm:bounds_sample_C_private}.  
\end{remark}
\section{Other Applications} \label{Sec:other_applications}

In this section, we show how the results presented in~\cref{Sec:contraction_QLDP} find use in several applications, including quantum fairness and learning settings, in which we provide formal guarantees on the level of fairness and stability of learning algorithms.

\subsection{Quantum Fairness through QLDP}

Classical decision models, including machine learning models, are prone to discriminating against individuals based on different characteristics, for example, skin color or gender~\cite{flores2016false}. This has even led to legal mandates of ensuring fairness. With the introduction of quantum machine learning models, there is also a risk of whether fairness will be ensured to both classical and quantum data fed into these algorithms. In~\cite{fairnessQ_verifying22}, it was shown that noisy quantum algorithms can improve fairness. 
Since fairness and privacy are related domains, it is vital to understand the impact of private algorithms on ensuring fairness as well.

Formally, quantum fairness aims to treat all input states equally, meaning that all pairs of input states that are close in some distance metric (e.g., close in normalized trace distance) should  yield similar outcomes when processed by a quantum channel~\cite{fairnessQ_verifying22}.
Define $\cA \coloneqq \cM \circ \cE$, which is a quantum-to-classical channel where a quantum channel~$\cE$ is followed by a measurement channel comprised of a POVM $\{M_i\}_{i \in \cO}$. With that, quantum fairness is defined in~\cite{fairnessQ_verifying22} as follows.

\begin{definition}[$(\alpha,\beta)$-Fairness~\cite{fairnessQ_verifying22}]
 Let $\cA = \cM \circ \cE$, and let $\hat{D}(\cdot \| \cdot)$ and $d(\cdot \| \cdot)$ be distance metrics on $\cD(\cH)$ and $\cD(\cO)$, respectively. Fix $0 < \alpha,\beta \leq 1$. Then the decision model $\cA$ is $(\alpha,\beta)$ fair if for all $\rho,\sigma \in \cD(\cH)$ such  that $\hat{D}(\rho \| \sigma) \leq \alpha$,
\begin{equation}
d\!\left( \cA(\rho)\| \cA(\sigma) \right) \leq \beta.
\end{equation}
\end{definition}

By choosing $\hat{D}$ to be the normalized trace distance and $d\!\left( \cA(\rho)\| \cA(\sigma) \right)= \frac{1}{2} \sum_i \!\left| \Tr\!\left[M_i \cE(\rho-\sigma) \right]\right |$, \cite[Proposition~14]{nuradha2023quantum} states that $\cA$ satisfying $\varepsilon$-QLDP implies that it is also $(\alpha, \sqrt{\varepsilon'/2})$-fair, where $\varepsilon'= \min\{\varepsilon,\varepsilon^2/2\}$. Note that for the case $\varepsilon >1$, the above bound could be weak.
With \cref{thm:contraction_coeff_TD}, we improve the existing bound on achievable fairness through $\varepsilon$-QLDP mechanisms, which is applicable for all $\varepsilon \geq 0 $, and we also extend them to cases for which $\delta \geq 0$.

\begin{proposition}[Fairness Guarantee from $(\varepsilon,\delta)$-QLDP]
\label{prop:Privacy-implying-fairness}
Suppose that $\hat{D}(\rho \| \sigma)= \frac{1}{2} \left \| \rho- \sigma \right\|_1 $ and $d\!\left( \cA(\rho)\| \cA(\sigma) \right)= \frac{1}{2} \sum_i \!\left| \Tr\!\left[M_i \cE(\rho-\sigma) \right]\right |$.
    If $\cE$ satisfies ($\varepsilon, \delta$)-QLDP, then $\cA=\{\cE, \{M_i\}_{i \in \cO} \}$ is $(\alpha, \varepsilon'(\alpha))$-fair for all $\rho,\sigma \in \cD(\cH)$ such  that $\hat{D}(\rho \| \sigma) \leq \alpha$, where 
    \begin{equation}
       \varepsilon'(\alpha) \coloneqq  \alpha \left( \frac{e^\varepsilon -1 + 2\delta}{e^\varepsilon +1} \right).
    \end{equation}
\end{proposition}
\begin{IEEEproof}
       From \cref{thm:contraction_T_eps_delta}, the channel $\cE$ being ($\varepsilon, \delta$)-QLDP implies that
       \begin{align}
          \frac{1}{2} \left\| \cE(\rho) - \cE(\sigma) \right\|_1 & \leq  \frac{e^\varepsilon -1 +2\delta}{e^\varepsilon +1} \left(\frac{1}{2}\left\| \rho - \sigma\right\|_1\right) \\ 
          &\leq \alpha  \frac{e^\varepsilon -1+2\delta}{e^\varepsilon +1},
       \end{align}
    where the last inequality follows from the assumption $\hat{D}(\rho \Vert \sigma) \leq \alpha$.
    
    Then, consider the measurement channel 
    that performs the following transformation:
    \begin{equation}
    \cE(\rho) \to \sum_{i \in \cO} \Tr\!\left[M_i \cE(\rho)\right] |i \rangle\!\langle i|.
    \end{equation}
    It follows from the data-processing inequality for the trace distance that 
    \begin{equation}
   \frac{1}{2} \left \| \sum_{i \in \cO} \left( \Tr\!\left[M_i \cE(\rho)\right]  -  \Tr\!\left[M_i \cE(\sigma)\right] \right) |i \rangle\!\langle i| \right \|_1 \leq \alpha  \frac{e^\varepsilon -1+2 \delta}{e^\varepsilon +1}  .
    \end{equation}
    This leads to 
    \begin{equation}
    d\!\left( \cA(\rho)\| \cA(\sigma) \right)= \frac{1}{2} \sum_i \left| \Tr\!\left[M_i \cE(\rho-\sigma) \right] \right | \leq \alpha  \frac{e^\varepsilon -1+2\delta}{e^\varepsilon +1} ,\end{equation}
    concluding the proof. 
\end{IEEEproof}

\subsection{Stability for Quantum Learning through Private Channels}

{Designing learning algorithms that also ensure privacy for input data is of importance, and it has been widely studied in the classical setting (for example, see \cite{kasiviswanathan2011can}).}
A learning algorithm is known to be stable if its output does not depend too much on any individual training data \cite{bousquet2002stability,raginsky2016information}. It is also known that stability and generalization of a classical learning algorithm to new inputs are closely related, where stability implies generalization using information-theoretic tools~\cite{raginsky2016information}.
Classical differentially private learners were proved to generalize well \cite[Theorem~5]{raginsky2016information}.
In~\cite[Proposition~6]{nuradha2022pufferfishJ}, it was shown that learners satisfying a mutual-information-based variant of differential privacy also satisfy algorithmic stability, and hence generalize well for new data. 

It is a natural research question to explore whether quantum private learners also provide stability and generalization in quantum learning settings. In~\cite{caro2023information}, it was shown that the generalization error of classical-quantum learners is bounded from above by a function of mutual information between input space and output space, and a Holevo information term (see~Theorem~1 therein for an example).
To this end, the authors of \cite{caro2023information}  show in Eqs.~(5.7.1)--(5.7.6) therein that the mutual information terms decay fast for learners comprised of $\varepsilon$-QLDP channels and left the analysis of the Holevo information term as an open question. In this section, we show  that $\varepsilon$-QLDP learners also provide Holevo information stability, and we do so by  bounding the Holevo information from above by a function of the QLDP privacy parameter~$\varepsilon$.

\medskip

\textbf{Holevo Information Stability from QLDP:}
Let $X \sim P_X$ be a random variable, which can take values in an alphabet $\cX$.
Depending on $X$, the state $\rho^X$ is chosen from the set $\{ \rho^1, \ldots, \rho^{|\cX|} \}$. Then the state $\rho^X$ is sent through a quantum channel $\cA_{A \to B}$ satisfying QLDP. Afterwards the goal is to identify $X$ by performing a measurement described by the POVM $\{M_y\}_{y\in \cY}$, which realizes the output $Y$. 
%The flow diagram relevant to this setup is shown in \cref{fig:classicalXY}.

Here, we focus on how much information about $X$ can be learned from the output of the quantum privacy mechanism $\cA(\rho^X)$
%and the classical output $Y$, 
with an emphasis on the Holevo information $\sI\!\left(X;B\right)_{\sigma}$. Here, we define the classical--quantum state
\begin{equation}
    \sigma_{XB} \coloneqq \sum_{x \in \cX} P_X(x) \, |x \rangle \! \langle x| \otimes \cA(\rho^x),
\end{equation}
and the Holevo information of  $\sigma_{XB}$ as
\begin{equation}
    \sI\!\left(X;B\right)_{\sigma} 
    \coloneqq  D(\sigma_{XB} \Vert \sigma_X \otimes \sigma_B) ,
\end{equation}
with $\sigma_X=\Tr_B[\sigma_{XB}]$ and $\sigma_B= \Tr_X[\sigma_{XB}]$.

 We show that Holevo information stability, i.e., $I(X;B) \leq \beta$, can be achieved by $\varepsilon$-QLDP mechanisms. To this end, we establish a bound improving upon the existing bound $ \beta = \min\{ \varepsilon , \varepsilon^2/2\}$ given in \cite[Proposition~15]{nuradha2023quantum}, with the improvement following because $\varepsilon \left( \frac{e^\varepsilon -1}{ e^\varepsilon +1}\right) \leq \min\{ \varepsilon , \varepsilon^2/2\}$ for all $\varepsilon \geq 0$.

\begin{proposition}[Holevo Information Stability from QLDP] \label{prop:Bounds on mutual information due to QLDP}
Let $\cA_{A \to B}$ be a quantum channel. If $\cA$ satisfies $\varepsilon$-QLDP, then the Holevo information has the following upper bound:
\begin{equation}
  \sI\!\left(X;B\right)_{\sigma} \leq \varepsilon \left( \frac{e^\varepsilon -1}{ e^\varepsilon +1}\right).  
\end{equation}
\end{proposition}

\begin{IEEEproof}
Consider that
\begin{align}
   \sI\!\left(X;B\right)_{\sigma} %\notag \\
  &= \sum_{x \in \cX} P_X(x) \,  D\!\left( \cA(\rho^x) \middle \Vert \sum_{x' \in \cX} P_X(x') \cA(\rho^{x'}) \right)   \\ 
  & \leq \sum_{x \in \cX} \sum_{x' \in \cX} P_X(x) P_X(x') \  D \!\left( \cA(\rho^x) \middle \Vert  \cA(\rho^{x'}) \right) \\
  & \leq \varepsilon \left( \frac{e^\varepsilon -1}{ e^\varepsilon +1}\right)\sum_{x \in \cX} \sum_{x' \in \cX} P_X(x) P_X(x') \  T(\rho_x, \rho_x') \\
  & \leq  \varepsilon \left( \frac{e^\varepsilon -1}{ e^\varepsilon +1}\right),
\end{align}
where the first inequality follows from the joint convexity of quantum relative entropy~\cite{LR73}. The second inequality follows from \cref{Cor:contraction_RE_TS} and the last inequality from $T(\rho,\sigma) \leq 1$ for states $\rho$ and $\sigma$, 
concluding the proof.
\end{IEEEproof}

\section{Concluding Remarks and Future Directions}

\label{Sec:conclusion}

In this work, we derived upper bounds on the contraction of quantum divergences under privacy constraints imposed by channels that satisfy $\varepsilon$-QLDP. Notably, we fully characterized the privatized contraction coefficient of trace distance, proving the converse by deriving an upper bound on the contraction of the hockey-stick divergence, and proving achievability by proposing a novel QLDP mechanism that achieves the bound. Next, with the tools developed, we studied quantum private hypothesis testing and provided bounds on the sample complexity, when we have access to privatized samples. We showcased the cost of privacy on this statistical task while providing tight sample complexity bounds on special sets of states, and special classes of private channels. In addition, we also characterized the privatized contraction coefficient of trace distance under $(\varepsilon,\delta)$-QLDP mechanisms. Finally, we explored how the contraction of divergences derived can be applied in other applications including ensuring fairness and providing formal guarantees on stability and generalization for quantum learning settings by addressing an open question posed in~\cite{caro2023information}. 

Future work includes exploring tight upper and lower bounds related to~\cref{thm:bounds_sample_C_private} in the general setting with $\delta>0$, also for the case where practically relevant classes of channels are considered; and analysing the cost of privacy and the impact of adaptive strategies on sample complexity. It is an interesting future direction to precisely characterize privatized contraction coefficients of other quantities (e.g., quantum relative entropy), similar to classically known results in \cite[Theorem~1]{Contraction_local_new24} for commuting states. We see that Bures distance has appeared in the analysis of a hypothesis testing task through~\cite{cheng2024sample}. To this end, finding contraction coefficients of the Bures distance, denoted as $\eta^\varepsilon_{B}$,  would potentially give a lower bound for the sample complexity as follows: for some constant $C>0$,
\begin{equation}
        \mathrm{SC}^{\varepsilon} _{(\rho,\sigma)} \geq C \frac{1}{\eta^\varepsilon_{B} \left[ d_{B}(\rho,\sigma)\right]  ^{2} }.
\end{equation}
Another interesting research question is to characterize the sample complexity of asymmetric binary quantum hypothesis testing. Here one could make use of the non-private sample-complexity bounds presented in~\cite[Theorem~9]{cheng2024sample}. Preliminary results in this direction are found in~\cref{rem:asym_hypoth_SC_lower}. {It is also an interesting future research direction to understand the impact of general noisy channels on the sample complexity of quantum hypothesis testing. Towards achieving this, main technical tool needed is the contraction coefficient of quantum divergences under general noisy channels.}

\section*{Acknowledgements} 
We thank Kaiyuan Ji for helpful discussions.
We are grateful to Hao-Chung Cheng, Christoph Hirche, and Cambyse Rouz\'e for coordinating the arXiv post of their paper titled "Sample Complexity of Locally Differentially Private Quantum Hypothesis Testing".
We would like to thank Behnoosh Zamanlooy and Shahab Asoodeh for sharing the full version with appendices of their conference paper~\cite{zamanlooy2023strong}, which inspired some of our proofs.
 TN  acknowledges support from the NSF under grant no.~2329662. MMW acknowledges support from the NSF under grants 2329662, 2315398, 2304816.

\bibliographystyle{IEEEtran}
\bibliography{reference}

\appendix

\subsection{Properties of Hockey-Stick Divergence} \label{app:properties_hocky_stick}

\begin{proposition}
Given a channel $\mathcal{A}$ and $\gamma\geq1$, the following equality holds:%
\begin{equation}
\sup_{\rho,\sigma}E_{\gamma}(\mathcal{A}(\rho)\Vert\mathcal{A}(\sigma
))=\sup_{\varphi_{1}\perp\varphi_{2}}E_{\gamma}(\mathcal{A}(\varphi_{1}%
)\Vert\mathcal{A}(\varphi_{2})),
\end{equation}
where the optimization on the left-hand side is over all states $\rho$ and
$\sigma$ and the optimization on the right-hand side is over orthogonal pure
states $\varphi_{1}$ and $\varphi_{2}$.
\end{proposition}

\begin{IEEEproof}
First, consider that the inequality
\begin{equation}
\sup_{\rho,\sigma}E_\gamma(\mathcal{A}(\rho)\Vert\mathcal{A}(\sigma))\geq
\sup_{\varphi_{1}\perp\varphi_{2}}E_\gamma(\mathcal{A}(\varphi_{1}%
)\Vert\mathcal{A}(\varphi_{2}))
\end{equation}
trivially holds due to the containment $\left\{  \left(  \varphi_{1}
,\varphi_{2}\right)  :\varphi_{1}\perp\varphi_{2}\right\}  \subset\left\{
\left(  \rho,\sigma\right)  \right\}  $. So it remains to prove the opposite
inequality:
\begin{equation}
\sup_{\rho,\sigma}E_\gamma(\mathcal{A}(\rho)\Vert\mathcal{A}(\sigma))\leq
\sup_{\varphi_{1}\perp\varphi_{2}}E_\gamma(\mathcal{A}(\varphi_{1}
)\Vert\mathcal{A}(\varphi_{2})).
\end{equation}
First, from the joint convexity of $E_\gamma$ (see \cite[Eq.~(II.17)]{hirche2023quantum}), it readily follows that
\begin{equation}
\sup_{\rho,\sigma}E_\gamma(\mathcal{A}(\rho)\Vert\mathcal{A}(\sigma))\leq
\sup_{\psi,\phi}E_\gamma(\mathcal{A}(\psi)\Vert\mathcal{A}(\phi)),
\end{equation}
so that it suffices to perform the optimization over pure states. As such, we
now prove that
\begin{equation}
\sup_{\psi,\phi}E_{\gamma}(\mathcal{A}(\psi)\Vert\mathcal{A}(\phi))\leq
\sup_{\varphi_{1}\perp\varphi_{2}}E_{\gamma}(\mathcal{A}(\varphi_{1}%
)\Vert\mathcal{A}(\varphi_{2})).
\label{eq:desired-ineq-final-pure-to-ortho}
\end{equation}
From \Cref{lem:op-ineq-E-gamma-pure-ortho}, the operator inequality in~\eqref{eq:key-op-ineq} holds.
Applying that then gives
\begin{align}
E_{\gamma}(\mathcal{A}(\psi)\Vert\mathcal{A}(\phi))  & =\sup_{M\geq0}\left\{
\operatorname{Tr}[M(\mathcal{A}(\psi)-\gamma\mathcal{A}(\phi))]:M\leq
I\right\}  \\
& \leq\sup_{M\geq0}\left\{  \operatorname{Tr}[M\lambda_{1}\left(
\mathcal{A}(\varphi_{1})-\gamma\mathcal{A}(\varphi_{2})\right)  )]:M\leq
I\right\}  \\
& \leq\sup_{M\geq0}\left\{  \operatorname{Tr}[M\left(  \mathcal{A}(\varphi
_{1})-\gamma\mathcal{A}(\varphi_{2})\right)  )]:M\leq I\right\}  \\
& =E_{\gamma}(\mathcal{A}(\varphi_{1})\Vert\mathcal{A}(\varphi_{2}))\\
& \leq\sup_{\varphi_{1}\perp\varphi_{2}}E_{\gamma}(\mathcal{A}(\varphi
_{1})\Vert\mathcal{A}(\varphi_{2})).
\end{align}
The first inequality follows from~\eqref{eq:key-op-ineq}. Since this inequality holds for all pure states $\psi$ and $\phi$, we conclude~\eqref{eq:desired-ineq-final-pure-to-ortho}. 
\end{IEEEproof}

\begin{lemma}
\label{lem:op-ineq-E-gamma-pure-ortho}
For pure states $\psi$ and $\phi$, a positive map $\mathcal{A}$, and
$\gamma\geq1$, the following operator inequality holds:%
\begin{equation}
\mathcal{A}(\psi)-\gamma\mathcal{A}(\phi)\leq\lambda_{1}\left[  \mathcal{A}%
(\varphi_{1})-\gamma\mathcal{A}(\varphi_{2})\right]  ,\label{eq:key-op-ineq}%
\end{equation}
where a spectral decomposition of $\psi-\gamma\phi$ is given by%
\begin{equation}
\psi-\gamma\phi=\lambda_{1}\varphi_{1}-\lambda_{2}\varphi_{2} \label{eq:spectral_decomposition_2_pure},
\end{equation}
with $\varphi_{1}$ and $\varphi_{2}$ orthogonal pure states and
\begin{align}
\lambda_{1}  & \coloneqq \frac{1}{2}\left[  \sqrt{\left(  \gamma+1\right)
^{2}-4\gamma F(\psi,\phi)}-\left(  \gamma-1\right)  \right]  \in\left[
0,1\right]  , \label{eq:eigen_values_with_Fidelity}\\
\lambda_{2}  & \coloneqq \frac{1}{2}\left[  \sqrt{\left(  \gamma+1\right)
^{2}-4\gamma F(\psi,\phi)}+\left(  \gamma-1\right)  \right]  \in\left[
\gamma-1,\gamma\right]  ,
\end{align}
and the fidelity $F(\psi,\phi)\coloneqq \left\vert \langle\psi|\phi\rangle\right\vert
^{2}$.
\end{lemma}
\begin{IEEEproof}
Consider $\gamma\geq1$ and the operator $\psi-\gamma\phi$. Now consider that
$|\psi\rangle$ and $|\phi\rangle$ span a two-dimensional subspace, and in this
subspace we can write%
\begin{equation}
|\psi\rangle=\cos(\theta)|\phi\rangle+\sin(\theta)|\phi^{\perp}\rangle,
\end{equation}
where $|\phi^{\perp}\rangle$ is a pure state vector orthogonal to
$|\phi\rangle$. Observe that $F(\psi,\phi)=\cos^{2}(\theta)$. Then%
\begin{equation}
|\psi\rangle\!\langle\psi|=\cos^{2}(\theta)|\phi\rangle\!\langle\phi|+\sin
(\theta)\cos(\theta)\left(  |\phi\rangle\!\langle\phi^{\perp}|+|\phi^{\perp
}\rangle\!\langle\phi|\right)  +\sin^{2}(\theta)|\phi^{\perp}\rangle\!\langle
\phi^{\perp}|
\end{equation}
and thus, in the basis $\left\{  |\phi\rangle,|\phi^{\perp}\rangle\right\}  $,%
\begin{align}
\psi-\gamma\phi &  =|\psi\rangle\!\langle\psi|-\gamma|\phi\rangle\!\langle\phi|\\
&  =%
\begin{bmatrix}
\cos^{2}(\theta)-\gamma & \sin(\theta)\cos(\theta)\\
\sin(\theta)\cos(\theta) & \sin^{2}(\theta)
\end{bmatrix}
.
\end{align}
For a $2\times2$ matrix $A$, the following well known expression exists for
its eigenvalues:%
\begin{equation}
\lambda_{\pm}=\frac{1}{2}\left(  \operatorname{Tr}[A]\pm\sqrt{\left(
\operatorname{Tr}[A]\right)  ^{2}-4\det(A)}\right)  .
\end{equation}
In this case, we find that $\operatorname{Tr}[\psi-\gamma\phi]=1-\gamma$ and
$\det(\psi-\gamma\phi)=-\gamma\sin^{2}(\theta)$, so that the eigenvalues of
$\psi-\gamma\phi$ are given by%
\begin{align}
\lambda_{\pm}  & =\frac{1}{2}\left(  1-\gamma\pm\sqrt{\left(  1-\gamma\right)
^{2}+4\gamma\sin^{2}(\theta)}\right)  \\
& =\frac{1}{2}\left(  1-\gamma\pm\sqrt{\left(  \gamma-1\right)  ^{2}%
+4\gamma\left(  1-F(\psi,\phi)\right)  }\right)  \\
& =\frac{1}{2}\left(  1-\gamma\pm\sqrt{\left(  \gamma+1\right)  ^{2}-4\gamma
F(\psi,\phi)}\right)  .
\end{align}
Thus, there exist orthogonal state vectors $|\varphi_{1}\rangle$ and
$|\varphi_{2}\rangle$ such that%
\begin{equation}
\psi-\gamma\phi=\lambda_{+}\varphi_{1}+\lambda_{-}\varphi_{2}.
\end{equation}
Observe that $\lambda_{+}\in\left[  0,1\right]  $ and $\lambda_{-}\in\left[
-\gamma,-\left(  \gamma-1\right)  \right]  $ because $F(\psi,\phi)\in\left[
0,1\right]  $. Now identifying $\lambda_{1}=\lambda_{+}$ and $\lambda
_{2}=-\lambda_{-}$, we can write%
\begin{equation}
\psi-\gamma\phi=\lambda_{1}\varphi_{1}-\lambda_{2}\varphi_{2},
\end{equation}
and conclude that $\lambda_{1}\in\left[  0,1\right]  $ and $\lambda_{2}%
\in\left[  \gamma-1,\gamma\right]  $, as claimed.

To prove the operator inequality in~\eqref{eq:key-op-ineq}, consider that%
\begin{align}
\mathcal{A}(\psi)-\gamma\mathcal{A}(\phi)  & =\mathcal{A}(\psi-\gamma\phi)\\
& =\mathcal{A}(\lambda_{1}\varphi_{1}-\lambda_{2}\varphi_{2})\\
& =\lambda_{1}\mathcal{A}(\varphi_{1})-\lambda_{2}\mathcal{A}(\varphi_{2})\\
& =\lambda_{1}\mathcal{A}(\varphi_{1})-\gamma\lambda_{1}\mathcal{A}%
(\varphi_{2})+\gamma\lambda_{1}\mathcal{A}(\varphi_{2})-\lambda_{2}%
\mathcal{A}(\varphi_{2})\\
& =\lambda_{1}\left[  \mathcal{A}(\varphi_{1})-\gamma\mathcal{A}(\varphi
_{2})\right]  +\left[  \gamma\lambda_{1}-\lambda_{2}\right]  \mathcal{A}%
(\varphi_{2})\\
& \leq\lambda_{1}\left[  \mathcal{A}(\varphi_{1})-\gamma\mathcal{A}%
(\varphi_{2})\right]  .
\end{align}
The last inequality follows because $\gamma\lambda_{1}-\lambda_{2}\leq0$ and
$\mathcal{A}$ is a positive map, so that $\left[  \gamma\lambda_{1}%
-\lambda_{2}\right]  \mathcal{A}(\varphi_{2})\leq0$. Indeed, consider that%
\begin{equation}
1-\gamma=\operatorname{Tr}[\psi-\gamma\phi]=\operatorname{Tr}[\lambda
_{1}\varphi_{1}-\lambda_{2}\varphi_{2}]=\lambda_{1}-\lambda_{2},
\end{equation}
which implies that%
\begin{align}
\gamma\lambda_{1}-\lambda_{2}  & =\gamma\lambda_{1}-\lambda_{1}+\lambda
_{1}-\lambda_{2}\\
& =\left(  \gamma-1\right)  \lambda_{1}+\left(  1-\gamma\right)  \\
& =-\left(  \gamma-1\right)  \left(  1-\lambda_{1}\right)  \\
& \leq0.
\end{align}
Here, we used the fact that $\gamma\geq1$ and $\lambda_{1}\in\left[
0,1\right]  $. 
\end{IEEEproof}

\begin{remark}[Hockey-Stick Divergence for Pure States]
 For two pure states $\psi$ and $\phi$ and $\gamma \geq 1$, their hockey-stick divergence  is equal to 
 \begin{equation}
       E_\gamma\!\left( \psi \Vert \phi \right)= \frac{1}{2} \left( \sqrt{(\gamma+1)^2 - 4 \gamma F(\psi, \phi) } + 1 - \gamma  \right).
 \end{equation}
 This follows by considering the positive eigenvalue in~\eqref{eq:spectral_decomposition_2_pure} together with~\eqref{eq:eigen_values_with_Fidelity}. Note that with the above equality, it follows that the upper bound in Eq.~(II.11) in~\cite{hirche2023quantum} is saturated for pure states.
\end{remark}

\subsection{Datta--Leditzky Divergence}\label{app:properties_datta_leditzky}

\begin{proposition}
Given a channel $\mathcal{A}$, $\gamma\geq1$, and $\delta \in [0,1)$, the following equality holds:%
\begin{equation}
\sup_{\rho,\sigma}\overline{\sD}^{\delta}(\mathcal{A}(\rho)\Vert\mathcal{A}(\sigma
))=\sup_{\varphi_{1}\perp\varphi_{2}}\overline{\sD}^{\delta}(\mathcal{A}(\varphi_{1}%
)\Vert\mathcal{A}(\varphi_{2})),
\end{equation}
where the optimization on the left-hand side is over all states $\rho$ and
$\sigma$ and the optimization on the right-hand side is over orthogonal pure
states $\varphi_{1}$ and $\varphi_{2}$.
\end{proposition}
\begin{IEEEproof}
First, consider that the inequality
\begin{equation}\label{eq:trivial_DL}
\sup_{\rho,\sigma}\overline{\sD}^{\delta}(\mathcal{A}(\rho)\Vert\mathcal{A}(\sigma))\geq
\sup_{\varphi_{1}\perp\varphi_{2}}\overline{\sD}^{\delta}(\mathcal{A}(\varphi_{1}%
)\Vert\mathcal{A}(\varphi_{2}))
\end{equation}
trivially holds due to the containment $\left\{  \left(  \varphi_{1}
,\varphi_{2}\right)  :\varphi_{1}\perp\varphi_{2}\right\}  \subset\left\{
\left(  \rho,\sigma\right)  \right\}  $. So it remains to prove the opposite
inequality:
\begin{equation}
\sup_{\rho,\sigma}\overline{\sD}^{\delta}(\mathcal{A}(\rho)\Vert\mathcal{A}(\sigma))\leq
\sup_{\varphi_{1}\perp\varphi_{2}}E_\gamma(\mathcal{A}(\varphi_{1}
)\Vert\mathcal{A}(\varphi_{2})).
\end{equation}
First, from the joint quasi-convexity of $\overline{\sD}^{\delta}$ (see \cite[Eq.~(67)]{nuradha2023quantum}), it readily follows that
\begin{equation}
\sup_{\rho,\sigma}\overline{\sD}^{\delta}(\mathcal{A}(\rho)\Vert\mathcal{A}(\sigma))\leq
\sup_{\psi,\phi}\overline{\sD}^{\delta}(\mathcal{A}(\psi)\Vert\mathcal{A}(\phi)),
\end{equation}
so that it suffices to perform the optimization over pure states. As such, we
now prove that
\begin{equation} 
\sup_{\psi,\phi}\overline{\sD}^{\delta}(\mathcal{A}(\psi)\Vert\mathcal{A}(\phi))\leq
\sup_{\varphi_{1}\perp\varphi_{2}}\overline{\sD}^{\delta}(\mathcal{A}(\varphi_{1}%
)\Vert\mathcal{A}(\varphi_{2})).
\label{eq:desired-ineq-final-pure-to-ortho_DL}
\end{equation}   
By~\cite[Corollary~1]{nuradha2023quantum}, we have that 
\begin{align}
   \sup_{\psi,\phi} \overline{\sD}^{\delta}\!\left(\cA(\psi)\Vert \cA(\phi) \right) & =\sup_{\psi,\phi} \  \ln \sup_{0 \leq W \leq I,  \Tr\left[W \cA(\psi)\right] \geq \delta} \frac{\Tr\!\left[W \cA(\psi) \right] - \delta }{\Tr\!\left[W \cA(\phi) \right]} \\ 
    &= \ln \sup_{\psi,\phi} \ \sup_{0 \leq W \leq I,  \Tr\left[ \cA^\dag(W) \psi\right] \geq \delta} \frac{\Tr\!\left[ \cA^\dag(W) \psi \right] - \delta }{\Tr\!\left[ \cA^\dag(W ) \phi \right]}.
    \end{align}
Then consider that 
\begin{align}
    \sup_{\psi,\phi} \overline{\sD}^{\delta}\!\left(\cA(\psi)\Vert \cA(\phi) \right) &=  \ln \sup_{\psi,\phi} \ \sup_{0 \leq W \leq I} \left\{ \frac{\Tr\!\left[ \cA^\dag(W) \psi \right] - \delta }{\Tr\!\left[ \cA^\dag(W ) \phi \right]}:   \Tr\left[ \cA^\dag(W) \psi\right] \geq \delta \right\} \\ 
    & \leq  \ln \sup_{\psi,\phi} \ \sup_{0 \leq W \leq I} \left\{ \frac{\Tr\!\left[ \cA^\dag(W) \psi \right] - \delta }{\Tr\!\left[ \cA^\dag(W ) \phi \right]}:   \lambda_{\max} \!\left(\cA^\dag(W)\right) \geq \delta \right\} \\ 
    &\leq \ln \sup_{0 \leq W \leq I} \left\{ \frac{\lambda_{\max} \!\left(\cA^\dag(W)\right) - \delta }{\lambda_{\min} \!\left(\cA^\dag(W)\right)}:   \lambda_{\max} \!\left(\cA^\dag(W)\right) \geq \delta \right\} \label{eq:upperbound_DL_achieved}
\end{align}
where the first inequality follows by the implication $\Tr\left[ \cA^\dag(W) \psi\right] \geq \delta \implies \lambda_{\max} \!\left(\cA^\dag(W)\right) \geq \delta $ with $\lambda_{\max}(B)$ denoting the maximum eigenvalue of $B$; the last inequality follows by $\lambda_{\min}(B) \leq \Tr[ B \psi] \leq \lambda_{\max}(B)$ for a state $\psi$ with $\lambda_{\min}(B)$ denoting the minimum eigenvalue of $B$.

Next, observe that the upper bound in~\eqref{eq:upperbound_DL_achieved} is achieved by picking $\psi$ and $\phi$ to be the pure states formed by the eigenvectors corresponding to the maximum and minimum eigenvalues of $\cA^\dag(W)$ for all $W$ such that $0\leq W \leq I$. Also note that these eigenvectors are orthogonal. With this, we conclude that~\eqref{eq:desired-ineq-final-pure-to-ortho_DL} holds and then together with~\eqref{eq:trivial_DL}, we conclude the proof of the proposition. 
%For a fixed $0 \leq W \leq I$ such that $\Tr\!\left[ \cA^\dag(W) \psi \right] \geq \delta$, observe that $\frac{\Tr\!\left[ \cA^\dag(W) \psi \right] - \delta }{\Tr\!\left[ \cA^\dag(W ) \phi \right]}$ is supremized by 
% $\psi = \lambda_{\max}\!\left( \cA^\dag(W)\right)$ and $\phi=\lambda_{\min}\!\left( \cA^\dag(W)\right)$, where $\lambda_{\max}(B)$ and $\lambda_{\min}(B)$ correspond to the pure states formed by the eigenvectors corresponding to the maximum eigenvalue of $B$ and the minimum eigenvalue of $B$, respectively. Also note that $\lambda_{\max}\!\left( \cA^\dag(W)\right)$ and $\lambda_{\min}\!\left( \cA^\dag(W)\right)$ are orthogonal pure states. Then for all $0 \leq W \leq I$ such that $\Tr\!\left[ \cA^\dag(W) \psi \right] \geq \delta$, the above reduction holds.
% With this, we conclude that~\eqref{eq:desired-ineq-final-pure-to-ortho_DL} holds and then together with~\eqref{eq:trivial_DL}, we conclude the proof of the proposition. 
\end{IEEEproof}

\subsection{Proof of~\cref{prop:tightness_f_contraction} } \label{App:proof_prop_f_contr_tight}
     \underline{Converse:}

    For all $\cA \in \cB^\varepsilon$ and $\rho$ and $\sigma$ states, by~\eqref{eq:contraction_f_d}, we have 
    \begin{align}
         \sup_{\substack{A \in \cB^\varepsilon,\\ \rho,\sigma \in \cD }}  D_f\!\left(\cA(\rho) \Vert \cA(\sigma) \right) &\leq   \sup_{\substack{A \in \cB^\varepsilon,\\ \rho,\sigma \in \cD }}  \frac{f\!\left( e^\varepsilon\right) + e^\varepsilon f\!\left( e^{-\varepsilon}\right) }{e^\varepsilon +1} T(\rho,\sigma) \\
        &\leq \frac{f\!\left( e^\varepsilon\right) + e^\varepsilon f\!\left( e^{-\varepsilon}\right) }{e^\varepsilon +1}, \label{eq:Converse_U}
    \end{align}
 where the last inequality follows because $T(\rho,\sigma) \leq 1$.  

 \medskip 
 \underline{Achievability:}
 First we define the measurement channel $\cM$ as 
 \begin{equation}\label{eq:measurement_channel_U}
        \cM(\omega) \coloneqq \Tr[ M \omega] |0 \rangle\!\langle 0| + \Tr[ (I-M) \omega] |1 \rangle\!\langle 1| ,
    \end{equation} 
where  $M$ is a measurement operator (satisfying $0  \leq M \leq I$) and $\omega$ is a quantum state. We also define the classical binary symmetric channel  $\cA^p_{\operatorname{BSC}}$ with the flip parameter $p \in [0,1/2]$ as follows:\begin{align}
    \cA^p_{\operatorname{BSC}}(|0\rangle\!\langle 0|) &= (1-p) |0\rangle\!\langle 0|+ p|1\rangle\!\langle 1|, \\
      \cA^p_{\operatorname{BSC}}(|1\rangle\!\langle 1| ) &= (1-p) |1\rangle\!\langle 1|+ p|0\rangle\!\langle 0|.
\end{align}
With that we consider the composite channel $\cA^p_{\operatorname{BSC}} \circ \cM$, and for the input state $\omega$ we have 
% where 
%  for a fixed measurement operator $M$ (satisfying $0  \leq M \leq I$) and for an input state $\omega$, the measurement channel $\cM$ is defined as 
%  \begin{equation}\label{eq:measurement_channel_U}
%         \cM(\omega) \coloneqq \Tr[ M \omega] |0 \rangle\!\langle 0| + \Tr[ (I-M) \omega] |1 \rangle\!\langle 1| 
%     \end{equation} 
% and 

% is the binary symmetric channel with the flip parameter $p \in [0,1/2]$ such that 

% With that, for state $\omega$ we have 
\begin{equation}
 \cA^p_{\operatorname{BSC}} \circ \cM (\omega)   = q_{0}^\omega |0 \rangle\!\langle 0| +  q_{1}^\omega |1 \rangle\!\langle 1|, 
\end{equation}
where 
\begin{align}
    q_{0}^\omega &\coloneqq (1-p) \Tr[M \omega] + p (1- \Tr[M\omega])  \\
    & = p + (1-2p)\Tr[M\omega] ,\\
    q_{1}^\omega & \coloneqq  p \Tr[M \omega] + (1-p) (1- \Tr[M\omega]) \\
    & = p + (1-2p)(1-\Tr[M\omega]).
\end{align}

For states $\rho$ and $\sigma$, we consider 
\begin{align} \label{eq:for_zero_bound}
    \frac{q_{0}^\rho}{q_{0}^\sigma} -1 & = \frac{q_{0}^\rho-q_{0}^\sigma}{q_{0}^\sigma} \\
    &= \frac{p + (1-2p) \Tr[M \rho] -(p +(1-2p) \Tr[M \sigma])}{p +(1-2p) \Tr[M \sigma]} \\ 
    & \leq \frac{(1-2p) \Tr[M(\rho-\sigma)]}{p} \\ 
    &\leq \frac{(1-2p)  T(\rho,\sigma)}{p}\\ 
    & \leq \frac{1-2p}{p},
\end{align}
where the first equality follows since $(1-2p) \Tr[M \sigma] \geq 0$ when $p \in[0,1/2]$; second inequality follows due to $T(\rho,\sigma)= \sup_{0\leq M \leq I} \Tr[M(\rho-\sigma)]$; and the last inequality by $T(\rho,\sigma) \leq 1$.

With that, we arrive at 
\begin{equation}
      \frac{q_{0}^\rho}{q_{0}^\sigma} \leq \frac{1}{p} -1 .
\end{equation}
Now, we choose $p = 1/(e^\varepsilon +1)$; then we see that 
\begin{equation}
    \frac{q_{0}^\rho}{q_{0}^\sigma} \leq \frac{1}{p} -1 = e^\varepsilon. 
\end{equation}

Similarly, we can show that 
\begin{equation}\label{eq:bound_e_var_0}
     \frac{q_{1}^\rho}{q_{1}^\sigma} \leq \frac{1}{p} -1 = e^\varepsilon
\end{equation}
by using the fact that 
\begin{equation}\label{eq:bound_e_var_1}
    q_1^\omega = p+ (1-2p) \Tr[(I-M) \omega] 
\end{equation}
and following analogous proof arguments as in~\eqref{eq:for_zero_bound}.

With that, we show the following: for all $\rho$ and $\sigma$ with $p= 1/(e^\varepsilon +1)$, 
\begin{align}
    E_{e^\varepsilon}\!\left(\cA^p_{\operatorname{BSC}} \circ \cM (\rho) \Vert \cA^p_{\operatorname{BSC}} \circ \cM(\sigma) \right) &= \Tr\!\left[ (\cA^p_{\operatorname{BSC}} \circ \cM (\rho) - e^\varepsilon \cA^p_{\operatorname{BSC}} \circ \cM(\sigma))_+ \right] \\ 
    &= \Tr\!\left[ \left((q_{0}^\rho-e^\varepsilon q_{0}^\sigma) |0\rangle\!\langle 0| +  (q_{1}^\rho-e^\varepsilon q_{1}^\sigma) |1\rangle\!\langle 1|\right)_{+}\right]\\ 
    &=\max\{ 0, q_{0}^\rho-e^\varepsilon q_{0}^\sigma\} \Tr[|0\rangle\!\langle 0|] +\max\{ 0, q_{1}^\rho-e^\varepsilon q_{1}^\sigma\} \Tr[|1\rangle\!\langle 1|] \\ 
    &= \max\{ 0, q_{0}^\rho-e^\varepsilon q_{0}^\sigma\} + \max\{ 0, q_{1}^\rho-e^\varepsilon q_{1}^\sigma\} \\ 
    &=0,
\end{align}
where the last equality follows by using~\eqref{eq:bound_e_var_0} and~\eqref{eq:bound_e_var_1}.

Then we conclude that
\begin{equation}
    \sup_{\rho,\sigma \in \cD} E_{e^\varepsilon}\!\left(\cA^p_{\operatorname{BSC}} \circ \cM (\rho) \Vert \cA^p_{\operatorname{BSC}} \circ \cM(\sigma) \right)=0,
\end{equation}
so that $\cA^p_{\operatorname{BSC}} \circ \cM \in \cB^\varepsilon$ satisfies $\varepsilon$-QLDP with $p=1/(e^\varepsilon +1)$ by~\eqref{eq:equivalent_HS_QLDP}.

Let us consider the special case in which $\rho'$ and $\sigma'$ are orthogonal states. Then choose $M= \Pi_{\rho'}$, which is the projection onto the support of the state $\rho'$. With that choice 
\begin{align}
    \Tr[M \rho'] =1 \quad \textnormal{and} 
    \quad  \Tr[M \sigma'] =0.
\end{align}

For states $\rho'$ and $\sigma'$, then we have 
\begin{align}
    q_0^{\rho'} &= 1-p, \\
    q_0^{\sigma'} &= p.
\end{align}
With that, we define two binary classical distributions as follows: $\omega \in \{\rho',\sigma'\}$
\begin{equation}
    q^\omega \coloneqq  q_0^{\omega} |0\rangle\!\langle 0| + (1- q_0^{\omega}) |1\rangle\!\langle 1|.
\end{equation}

Recall that for classical discrete distributions $p$ and $q$, $f$-divergences can be written as 
\begin{align}
   D_f(p\Vert q) &\coloneqq \sum_{x} q(x) f\!\left( \frac{p(x)}{q(x)} \right)  \\
   &= \int_{1}^{\infty} f''(\gamma) E_\gamma(p \Vert q) + \gamma^{-3} f''(\gamma^{-1}) E_\gamma(q \Vert p) \dd \gamma.
\end{align}
For this setting, we have
\begin{align}
    D_f\!\left(\cA^p_{\operatorname{BSC}} \circ \cM (\rho') \Vert \cA^p_{\operatorname{BSC}} \circ \cM(\sigma') \right) &= D_f(q^{\rho'} \Vert q^{\sigma'}) \\
    &= \sum_{x\in \{0,1\}} q^{\sigma'}_x f\!\left( \frac{q^{\rho'}_x }{q^{\sigma'}_x } \right) \\
    &= p f\!\left( \frac{ 1-p }{p} \right) + (1-p) f\!\left( \frac{ p }{1-p} \right) \\
    &= \frac{1}{e^\varepsilon+1} f(e^\varepsilon) + \frac{e^\varepsilon}{e^\varepsilon +1}f (e^{-\varepsilon}),
\end{align}
where the last equality follows by substituting $p=1/(e^\varepsilon +1)$.

This then leads to 
\begin{align}\label{eq:achievability_U}
    \frac{1}{e^\varepsilon+1} f(e^\varepsilon) + \frac{e^\varepsilon}{e^\varepsilon +1}f (e^{-\varepsilon}) = D_f\!\left(\cA^p_{\operatorname{BSC}} \circ \cM (\rho') \Vert \cA^p_{\operatorname{BSC}} \circ \cM(\sigma') \right) \leq \sup_{\substack{A \in \cB^\varepsilon,\\ \rho,\sigma \in \cD }}   D_f\!\left(\cA(\rho) \Vert \cA(\sigma) \right),
\end{align}
by recalling that $\cA^p_{\operatorname{BSC}} \circ \cM  \in \cB^\varepsilon$.

Finally, we conclude the proof by combining~\eqref{eq:Converse_U} and~\eqref{eq:achievability_U}.

\subsection{Sample Complexity in the Low-Privacy Regime}

\label{app:SC_low}

In this appendix, we prove~\cref{prop:SC_low_privacy}.
We use the following shorthand:
\begin{equation}
    d_B^2(\rho,\sigma) \coloneqq \left[d_B(\rho,\sigma)\right]^2.
\end{equation}
Let $\rho$ and $\sigma$ be states with dimension $d$.
Let $\cM$ be the measurement channel comprised of the POVM formed by the eigenbasis of $\rho \# \sigma^{-1}$ with $k \leq d$ POVM elements,
where $A \# B$ denotes the matrix geometric mean of positive semi-definite operators $A$ and $B$. It is defined for $A$ and $B$ positive definite operators as
\begin{equation}
    A \# B \coloneqq A^{1/2}\left( A^{-1/2} B A^{-1/2} \right)^{1/2} A^{1/2}. \label{eq:geometric-mean}
\end{equation}
and as $\lim_{\varepsilon \to 0^+} (A+\varepsilon I)\#(B+\varepsilon I)$ in the more general case when $A$ and $B$ are positive semi-definite.
With that measurement (POVM) achieving fidelity~\cite{fuchs1995mathematical} and recalling~\eqref{eq:Bures}, we have 
\begin{equation}\label{eq:equal_Bures_M}
    d_B^2(\rho,\sigma)= d_B^2\!\left( \cM(\rho), \cM(\sigma) \right).
\end{equation}
Also recall that 
\begin{equation}
   \left[ T(\rho,\sigma) \right]^2 \leq d_B^2(\rho,\sigma) \leq 2 T(\rho,\sigma). 
\end{equation}

Fix $p=2/(e^\varepsilon +1)$ and $d=2$. Also choose $\cA \coloneqq \cA^p_{\mathrm{Dep}} \circ \cM \in \cB^\varepsilon$. 
With these in hand, consider that
\begin{align}
    d_B^2\!\left( \cA(\rho), \cA(\sigma) \right) & \geq  \left[ T\!\left(\cA(\rho),\cA(\sigma)\right) \right]^2 \label{eq:d_2_start} \\
    &= (1-p)^2 \left[ T\!\left(\cM(\rho),\cM(\sigma)\right) \right]^2 \\
    & \geq  \frac{(1-p)^2}{4} \left[d_B^2\!\left( \cM(\rho), \cM(\sigma) \right)\right]^2 \\ 
    & \geq \frac{1}{4} d_B^2\!\left( \cM(\rho), \cM(\sigma) \right) \min\! \left\{1,   (1-p)^2 d_B^2\!\left( \cM(\rho), \cM(\sigma) \right) \right\} \label{eq:d_2_end}
\\ 
&= \frac{1}{4} d_B^2\!\left( \rho, \sigma \right) \min\! \left\{1,   (1-p)^2 d_B^2\!\left( \rho, \sigma \right) \right\},
\end{align}
where the last equality follows from~\eqref{eq:equal_Bures_M}.

This leads to the inequality
\begin{equation}
     d_B^2\!\left( \cA(\rho), \cA(\sigma) \right) \geq \frac{1}{4} d_B^2\!\left( \rho, \sigma \right) 
\end{equation}
if 
\begin{equation}
    (1-p)^2 \geq \frac{1}{d_B^2(\rho,\sigma)}.
\end{equation}

By data processing, for all $\cA \in \cB^\varepsilon$, we have 
\begin{equation}
    d_B^2\!\left( \cA(\rho), \cA(\sigma) \right) \leq d_B^2(\rho,\sigma) .
\end{equation}
Combining both these bounds, we have 
\begin{equation}
    \frac{1}{ d_B^2(\rho,\sigma)} \leq \inf_{\cA \in \cB^\varepsilon} \frac{1}{d_B^2\!\left( \cA(\rho), \cA(\sigma) \right)} \leq \frac{4}{d_B^2(\rho,\sigma)}
\end{equation}
if 
\begin{equation}
    \left( \frac{e^\varepsilon -1}{e^\varepsilon +1}\right)^2 \geq \frac{1}{d_B^2(\rho,\sigma)}. 
\end{equation}
Then under this constraint on $\varepsilon$, together with~\eqref{eq:SC_no_privacy_d_B} and~\eqref{eq:SC_private_int}   we have that
\begin{equation}
     {\mathrm{SC}}^{\varepsilon} _{(\rho,\sigma)}=\Theta\!\left(  \frac{1}{ d_B^2(\rho,\sigma)} \right).
\end{equation}

Next, we consider the case where $\rho$ and $\sigma$ are general quantum states. Recall that $k$ is the number of POVM elements formed by the eigenbasis of $\rho \# \sigma^{-1}$. Let $\cN$ be a classical channel with input dimension~$k$ and output dimension two.
By applying~\cite[Corollary~3.4]{Varun_communication_cons_SDPI} and considering that $\cM'= \cN \circ \cM$  is a two-outcome measurement channel, we have that 
\begin{equation}
    1 \leq \frac{ d_B^2\!\left( \cM(\rho), \cM(\sigma) \right) }{ d_B^2\!\left( \cM'(\rho), \cM'(\sigma) \right) } \leq  1800 \max\!\left\{1, \frac{\min\{k,k'\}}{2} \right\}, 
\end{equation}
where $k' \coloneqq \ln\!\left(4/d_B^2(\rho,\sigma)\right)$.

 With the choice $\cA' \coloneqq \cA^p_{\mathrm{Dep}} \circ \cM'
$ and $p=2/(e^\varepsilon +1)$ following a similar approach as done from~\eqref{eq:d_2_start}-\eqref{eq:d_2_end}, we find that
\begin{align}
    d_B^2\!\left( \cA'(\rho), \cA'(\sigma) \right) & \geq  \left[ T\!\left(\cA'(\rho),\cA'(\sigma)\right) \right]^2 \\
    &= (1-p)^2 \left[ T\!\left(\cM'(\rho),\cM'(\sigma)\right) \right]^2 \\
    & \geq  \frac{(1-p)^2}{4} \left[d_B^2\!\left( \cM'(\rho), \cM'(\sigma) \right)\right]^2 \\ 
    &  \geq  \frac{(1-p)^2}{4 (1800)^2 L_{k,k'}} \left[d_B^2\!\left( \rho, \sigma \right)\right]^2 \\ 
    & \geq \frac{1}{4 (1800)^2 L_{k,k'}} d_B^2\!\left( \rho, \sigma \right) \min\! \left\{1,   (1-p)^2 d_B^2\!\left( \rho, \sigma \right) \right\},
\end{align}
where $L_{k,k'}$ is defined in~\eqref{eq:L_k_k_'}.

If $(1-p)^2 \geq 1/ d_B^2(\rho,\sigma)$, then we have that
\begin{equation}
     d_B^2\!\left( \cA'(\rho), \cA'(\sigma) \right) \geq \frac{1}{4 (1800)^2 L_{k,k'}} d_B^2\!\left( \rho, \sigma \right).
\end{equation}
This leads to 
\begin{equation}
     \frac{1}{ d_B^2(\rho,\sigma)} \leq \inf_{\cA \in \cB^\varepsilon} \frac{1}{d_B^2\!\left( \cA(\rho), \cA(\sigma) \right)} \leq \frac{4  (1800)^2 L_{k,k'} }{d_B^2(\rho,\sigma)}.
\end{equation}
Again together with~\eqref{eq:SC_no_privacy_d_B} and~\eqref{eq:SC_private_int}, we conclude the proof.

\subsection{Proof of the Lower Bound in \cref{rem:different_private_mechanisms}} \label{APP:proof_different_mechanisms}

In this appendix, we prove the lower bound given in~\eqref{eq:lower_bound_different_mechanism}. 

First we prove the following: 
\begin{equation}
p_{e}(\rho,\sigma,p,q)  \geq 2 \ \min\{p,q\} \ p_{e}(\rho,\sigma,\sfrac{1}{2},\sfrac{1}{2})     
\end{equation}
To obtain that, consider the following chain of relations: 
\begin{align}
   & p_{e}(\rho,\sigma,p,q)  \cr 
   &= \min_{\substack{M_{1},M_{2}\geq0: \\ M_1 + M_2 = I}}p\operatorname{Tr}%
[M_{2}\rho]+q\operatorname{Tr}[M_{1}\sigma] \\ 
& \geq 2 \min\{p,q\} \min_{\substack{M_{1},M_{2}\geq0: \\ M_1 + M_2 = I}} \frac{1}{2} \operatorname{Tr}%
[M_{2}\rho]+ \frac{1}{2}\operatorname{Tr}[M_{1}\sigma] \\
&=  2 \min\{p,q\}  \ p_{e}(\rho,\sigma,\sfrac{1}{2},\sfrac{1}{2}).
\end{align}
Let the fixed error probability be $\alpha$. Then, in the private setting choose $\cA_i \in \cB^\varepsilon$ for all $i \in\{1,\ldots,n\}$. With that we have 
\begin{align}
    \alpha & \geq p_{e}\!\left( \cA_1(\rho) \otimes \cdots \otimes \cA_n(\rho),\cA_1(\sigma) \otimes \cdots \otimes \cA_n(\sigma),p,q \right) \\
    &   \geq  2 \min\{p,q\} \ p_{e}\!\left(\cA_1(\rho) \otimes \cdots \otimes \cA_n(\rho),\cA_1(\sigma) \otimes \cdots \otimes \cA_n(\sigma),\sfrac{1}{2},\sfrac{1}{2} \right) \\
    &=  \min\{p,q\} \left(  1- \frac{1}{2}\left\Vert \cA_1(\rho) \otimes \cdots \otimes \cA_n(\rho) - \cA_1(\sigma) \otimes \cdots \otimes \cA_n(\sigma) \right\Vert _{1}\right). 
\end{align}
Rearranging the last inequality yields the following: 
\begin{equation}
    T\!\left( \cA_1(\rho) \otimes \cdots \otimes \cA_n(\rho),\cA_1(\sigma) \otimes \cdots \otimes \cA_n(\sigma)\right) \geq 1- \frac{\alpha}{ \min\{p,q\}} \eqqcolon \beta.
\end{equation}
Applying Pinsker's inequality, we get 
\begin{multline}\label{eq:applying_pinsker}
    D\!\left( \cA_1(\rho) \otimes \cdots \otimes \cA_n(\rho) \Vert \cA_1(\sigma) \otimes \cdots \otimes \cA_n(\sigma)\right)  \\
    \geq 2 \left[  T\!\left( \cA_1(\rho) \otimes \cdots \otimes \cA_n(\rho),\cA_1(\sigma) \otimes \cdots \otimes \cA_n(\sigma)\right)\right]^2.
\end{multline}
Together with the former, consider that
\begin{align}
  2  \beta^2 & \leq   D\!\left( \cA_1(\rho) \otimes \cdots \otimes \cA_n(\rho) \Vert \cA_1(\sigma) \otimes \cdots \otimes \cA_n(\sigma)\right) \\ 
    & = \sum_{i=1}^n \ D\!\left( \cA_i(\rho) \Vert \cA_i(\sigma)\right)  \label{eq:lower_bound_to_next}\\ 
    & \leq \sum_{i=1}^n \varepsilon \frac{e^\varepsilon -1}{e^\varepsilon +1 } T(\rho, \sigma)  \\ 
    & \leq n \varepsilon \frac{e^\varepsilon -1}{e^\varepsilon +1 } T(\rho, \sigma),
\end{align}
where the first equality follows from the additivity of quantum relative entropy and the first inequality follows from \cref{Cor:contraction_RE_TS} with the assumption $\cA \in \cB^\varepsilon$.

Thus, by rearranging the terms we conclude the proof of the lower bound in~\eqref{eq:lower_bound_different_mechanism}. 

 \subsection{Proof of \cref{thm:instance_sample_C_private}} 
 \label{App:Proof_instance_SC}

 \underline{Upper bound:}
The upper bound from \cref{thm:bounds_sample_C_private} still holds in this special case.
The only aspect remaining to show here is that 
$ \cA^p_{\mathrm{Dep}} \circ \cM\in \mathcal{W}^{\varepsilon}_{(\rho,\sigma)}$ when $p=2/(e^\varepsilon+1)$ where $\cM$ is as defined in~\eqref{eq:measurement_channel} with the choice $M$  to be the positive eigenspace of $\rho -\sigma$ and $\cA^p_{\mathrm{Dep}}$ is a depolarizing channel with $p>0$, which establishes that $\lambda_{\min}\!\left( \cA^p_{\mathrm{Dep}}(\omega) \right)\geq p/d >0$ for every state $\omega$. Putting everything together, we have $\lambda_{\min}\!\left( \cA^p_{\mathrm{Dep}}\circ \cM(\omega) \right)\geq 1/(e^\varepsilon +1) >0$ by recalling that $d=2$ is the output space of $\cM$.
Arguing the above, concludes the proof of the upper bound in~\eqref{eq:instance-specific-SC-bnds}. 

\underline{Lower bound}:
We start from~\eqref{eq:SC_to_RE} in the proof of \cref{thm:bounds_sample_C_private}. Assume that 
$\max\{\lambda_{\min}  \left( \cA(\rho) \right),\lambda_{\min}  \left( \cA(\sigma) \right) \}= \lambda_{\min}  \left( \cA(\sigma) \right)$ since $\cA \in \mathcal{W}^{\varepsilon}_{(\rho,\sigma)}$ we have $\lambda_{\min}  \left( \cA(\sigma) \right) >0$. 
Consider
\begin{align}
   D\!\left( \cA(\rho) \Vert \cA(\sigma)\right) 
    & \leq    \frac{4 \left(T\!\left( \cA(\rho), \cA(\sigma) \right) \right)^2}{\lambda_{\min}  \left( \cA(\sigma) \right) } \\
    & \leq \frac{ 4}{\lambda_{\min}  \left( \cA(\sigma) \right)}  \left(\frac{e^\varepsilon -1}{e^\varepsilon +1 } \right)^2 \left(T(\rho,\sigma)\right)^2 \label{eq:last_inequality}
\end{align}
where the first inequality follows by~\cite[Theorem~2]{audenaert2005continuity} and the monotonicity of norms such that $\|A\|_2 \leq \|A\|_1$, and the last inequality by applying \cref{thm:contraction_coeff_TD} since $\cA$ satisfies $\varepsilon$-QLDP.

Plugging the above into~\eqref{eq:SC_to_RE}, we obtain that 
\begin{align}
\mathrm{SC}^{\varepsilon}_{(\rho,\sigma)} & \geq \inf_{\cA \in \mathcal{W}^{\varepsilon}_{(\rho,\sigma)}} \frac{\lambda_{\min}  \left( \cA(\sigma) \right)}{4}  \ln\!\left( \frac{pq} {\alpha}\right) \left( \frac{(e^\varepsilon +1)}{(e^\varepsilon -1) T(\rho,\sigma)} \right)^2 \\ 
&\geq \frac{1}{e^\varepsilon +1} \ln\!\left( \frac{pq} {\alpha}\right) \left( \frac{(e^\varepsilon +1)}{(e^\varepsilon -1) T(\rho,\sigma)} \right)^2, 
\end{align}
where the last inequality follows from the assumption in the set~\eqref{eq:special_set_channels} that ensures that %there exist $C >0$ such that 
\begin{equation}
  \inf_{\cA \in \mathcal{W}^{\varepsilon}_{(\rho,\sigma)}}\frac{\max\{\lambda_{\min}  \left( \cA(\sigma) \right),\lambda_{\min}  \left( \cA(\rho) \right)\}}{4} = \frac{1}{e^\varepsilon +1} >0. 
\end{equation}

For the last part of the proof, consider that for $\varepsilon <1$, from~\eqref{eq:upper_highE} and~\eqref{eq:lower_highE} we have 
\begin{equation}
    \frac{1}{\varepsilon} \leq \frac{e^\varepsilon +1}{e^\varepsilon -1} 
        \leq \frac{4}{\varepsilon}.
\end{equation}
Then, together with the fact $e^\varepsilon +1 \leq e +1$ for $\varepsilon <1$, we conclude the proof.

\subsection{Other Variants of QLDP and Relation to $\varepsilon$-QLDP} \label{app:other_connections_eps_delta}

In the main text, we discussed the impact of privacy imposed by $\varepsilon$-QLDP with $\delta=0$ on the sample complexity of private quantum hypothesis testing. Next, we show several relations between $\varepsilon'$-QLDP and $(\varepsilon,\delta)$-QLDP framework when $\delta >0$, so that we may use the derived results from ~\cref{Sec:private_hypothesis_testing} to infer about the impact of privacy imposed by $(\varepsilon,\delta)$-QLDP mechanisms.

\medskip 
From~\cite[Lemma~5]{nuradha2023quantum}, we have that $(\varepsilon+\delta,0)$-QLDP implies that $(\varepsilon, \delta')$-QLDP, where 
\begin{equation}
    \delta' \coloneqq 1- \frac{e^{\varepsilon} +1}{e^{\varepsilon+\delta}+1}. 
\end{equation}
However, in this case $\delta'$ is dependent on $\varepsilon$. Next, we show another relation between the two frameworks without this dependence. 
\begin{proposition} \label{prop:implication_eps_delta}
Every $\cA$ that satisfies $ (\varepsilon+\delta,0)$-QLDP also satisfies $(\varepsilon,\delta)${-QLDP}. i.e.,
    \begin{equation}
        (\varepsilon+\delta,0)\mathrm{-QLDP} \ \quad \implies \quad  (\varepsilon,\delta)\mathrm{-QLDP}.
    \end{equation}
\end{proposition}
\begin{IEEEproof} Proof follows analogous to the proof of one implication of~\cite[Lemma~5]{DP_testing18} established for classical differential privacy.
Fix $ 0\leq M \leq I$ and assume that $\cA$ satisfies $(\varepsilon,\delta)$-QLDP.
    Then, consider 
    \begin{align}
        \Tr\!\left[M \cA(\rho) \right] & \leq e^{\varepsilon + \delta}  \Tr\!\left[M \cA(\sigma) \right] \\
        &= e^\varepsilon e^\delta \Tr\!\left[M \cA(\sigma) \right] + e^\varepsilon \Tr\!\left[M \cA(\sigma) \right] -e^\varepsilon \Tr\!\left[M \cA(\sigma) \right] \\
        &= e^\varepsilon \Tr\!\left[M \cA(\sigma) \right]+ (e^\delta -1) e^\varepsilon \Tr\!\left[M \cA(\sigma) \right].
    \end{align}
If $e^\varepsilon \Tr\!\left[M \cA(\sigma) \right] + \delta \geq 1$, then the following holds:
\begin{equation}
     \Tr\!\left[M \cA(\rho) \right] \leq 1 \leq e^\varepsilon \Tr\!\left[M \cA(\sigma) \right] + \delta.
\end{equation}
Now assuming $e^\varepsilon \Tr\!\left[M \cA(\sigma) \right] + \delta <1$, we have $e^\varepsilon \Tr\!\left[M \cA(\sigma) \right] < (1-\delta)$. With this, consider 
\begin{align}
    (e^\delta -1) e^\varepsilon \Tr\!\left[M \cA(\sigma) \right] & \leq  (e^\delta -1)(1-\delta) \\
    & \leq (e^\delta -1) e^{-\delta} \\
    &= 1-e^{-\delta} \\
    & \leq \delta,
\end{align}
where the second and the last inequality follows from $1-x \leq e^{-x}$ for all $x \in \RR$.
\end{IEEEproof}

With the above implication together with~\cref{thm:bounds_sample_C_private}, we have 
\begin{equation}
     \mathrm{SC}^{\varepsilon,\delta} _{(\rho,\sigma)} \coloneqq \inf_{\cA \in \cB^{\varepsilon,\delta}} \mathrm{SC}^{\cA} _{(\rho,\sigma)}\leq \inf_{\cA \in \cB^{\varepsilon+ \delta}} \mathrm{SC}^{\cA} _{(\rho,\sigma)} \leq  \left \lceil 2 \ln\!\left( \frac{\sqrt{pq}}{\alpha} \right) \!\left( \frac{(e^{\varepsilon+\delta} +1)}{(e^{\varepsilon+\delta} -1) T(\rho, \sigma)} \right)^2  \right\rceil,
\end{equation}
where $\cB^{\varepsilon,\delta}$ is the set of all $(\varepsilon,\delta)$-QLDP mechanisms. First inequality there follows from $(\varepsilon+\delta,0)$-QLDP mechanisms are $(\varepsilon,\delta)$-QLDP from~\cref{prop:implication_eps_delta}, which makes the latter a larger set of mechanisms. The second inequality follows from the upper bound in~\cref{thm:bounds_sample_C_private}.

\bigskip
Inspired by \cite[Lemma~3.2]{Lemma_DP_C} for classical differential privacy, next we show that $(\varepsilon,\delta)$-QLDP mechanism also implies $(\varepsilon',0)$-QLDP for another mechanism that depends on the original mechanism.
\begin{proposition} \label{prop:eps_delta_equvi_eps_prime}
    Let $\cA$ satisfy $(\varepsilon,\delta)$-QLDP. Then, there exists $\cA'$ such that $T\!\left( \cA(\omega), \cA'(\omega)\right) \leq \eta$ for $\eta \in [0,1]$ 
    $\cA'$ %all $\omega \in \cD(\cH)$ that
    satisfies $(\varepsilon',0)$-QLDP, where 
    \begin{equation}
        \varepsilon' \coloneqq \varepsilon + \ln\!\left(1+ \frac{d \delta}{\eta}e^{-\varepsilon} \right) 
        % \leq  \varepsilon + \frac{d \delta}{\eta} e^{-\varepsilon}
    \end{equation}
    with $d$ being the dimension of the Hilbert space $\cH$.
\end{proposition}
\begin{IEEEproof}
    Choose $\cA' = \cA_{\mathrm{Dep}}^\eta \circ \cA$ such that 
    \begin{equation}
        \cA'(\omega) = (1-\eta) \cA(\omega) + \eta \frac{I}{d},
    \end{equation}
    for $\eta \in (0,1)$.
For that choice, 
\begin{align}
    T\!\left( \cA(\omega), \cA'(\omega)\right) & =\eta T\!\left( \cA(\omega), \frac{I}{d}\right) \\
    &\leq \eta,
\end{align}
where the last inequality follows from the normalized trace distance being bounded from above by one.

Since $\cA$ satisfies $(\varepsilon,\delta)$-QLDP and due to data processing property of QLDP mechanisms, $\cA'$  also satisfies $(\varepsilon,\delta)$-QLDP. This leads to the following inequality for all $0 \leq M \leq I$ and states $\rho$, $\sigma$:
\begin{equation}
    \Tr\!\left[ M \cA'(\rho)\right] \leq e^\varepsilon   \Tr\!\left[ M \cA'(\sigma)\right] + \delta.
\end{equation}

By choosing rank one projection $P$, we obtain the above inequality by replacing $M$ with $P$. 
For that choice we have 
\begin{align}
      \Tr\!\left[ P \cA'(\sigma)\right] &=   \Tr\!\left[ P \left((1-\eta) \cA(\sigma) + \eta \frac{I}{d} \right)\right] \\ 
      &= (1-\eta)  \Tr\!\left[ P \cA(\sigma)\right]+ \frac{\eta \Tr[P]}{d} \\
      &\geq \frac{\eta }{d}, \label{eq:lower_bound_projection_prob}
\end{align}
where the last inequality follows from $(1-\eta)  \Tr\!\left[ P \cA(\sigma)\right] \geq 0$ and $\Tr[P]=1$. 

Now, let us consider 
\begin{align}
     \Tr\!\left[ P \cA'(\rho)\right] & \leq e^\varepsilon   \Tr\!\left[ P \cA'(\sigma)\right] + \delta \\ 
     &=  e^\varepsilon   \Tr\!\left[ P \cA'(\sigma)\right] + \frac{d\delta }{\eta} \times \frac{\eta}{d} \\
     & \leq e^\varepsilon   \Tr\!\left[ P \cA'(\sigma)\right] + \frac{d\delta }{\eta}  \Tr\!\left[ P \cA'(\sigma)\right] \\
     &= \left( e^\varepsilon + \frac{d\delta }{\eta}  \right) \Tr\!\left[ P \cA'(\sigma)\right],
\end{align}
where the penultimate inequality follows from~\eqref{eq:lower_bound_projection_prob}.

The above holds for all rank one projections. For $M=0$, the above inequality holds trivially.
We can write a general measurement $0 < M \leq I$ using spectral theorem as $M= \sum_i \lambda_i P_i$ with $\lambda_i > 0$ and $P_i$ for all $i$ are rank one projections. 
Having $\Tr\!\left[ P_i \cA'(\rho)\right] \leq \left( e^\varepsilon + \frac{d\delta }{\eta}  \right) \Tr\!\left[ P_i \cA'(\sigma)\right]$ for all $i$, then multiplying both sides by $\lambda_i$, and summing over all $i$, we arrive at 
\begin{equation}
    \sum_i \lambda_i \Tr\!\left[ P_i \cA'(\rho)\right] \leq \left( e^\varepsilon + \frac{d\delta }{\eta}  \right) \sum_i \lambda_i \Tr\!\left[ P_i \cA'(\sigma)\right].
\end{equation}
This is equivalent to 
\begin{equation}
    \Tr\!\left[ M \cA'(\rho)\right] \leq \left( e^\varepsilon + \frac{d\delta }{\eta}  \right) \Tr\!\left[ M \cA'(\sigma)\right],
\end{equation}
proving that $\cA'$ satisfies $(\varepsilon',0)$-QLDP with 
\begin{equation}
    \varepsilon'\coloneqq \ln\!\left(e^\varepsilon + \frac{d\delta }{\eta} \right).
\end{equation}
Simplifying the above with the identity $\ln(ab)= \ln(a) +\ln(b)$ for positive $a$ and $b$, provides the desired result concluding the proof.
\end{IEEEproof}

\medskip 
However, it is not quite clear on how to use~\cref{prop:eps_delta_equvi_eps_prime} directly to obtain a lower bound on the sample complexity of quantum hypothesis testing under $(\varepsilon,\delta)$-QLDP. We leave this as an open question.

\end{document}